\documentclass[%
 reprint, 
superscriptaddress,
preprintnumbers,
 amsmath,amssymb,
 aps, 
 prc,
floatfix,
]{revtex4-2}
\usepackage[table,dvipsnames]{xcolor}
\definecolor{VividPurple}{HTML}{9B00FF} 
\definecolor{VividMagenta}{HTML}{FF00AA}
\definecolor{VividBlue}{HTML}{005CFF}
\definecolor{VividOrange}{HTML}{FF5500}
\definecolor{VividRed}{HTML}{FF0033}      
\definecolor{VividGreen}{HTML}{00CC22}    
\definecolor{VividCyan}{HTML}{00E5FF}     
\definecolor{VividYellow}{HTML}{FFEE00}   
\definecolor{VividTeal}{HTML}{009FFF}     
\definecolor{VividPink}{HTML}{FF0088}     
\definecolor{VividIndigo}{HTML}{4B00FF}   
\definecolor{VividLime}{HTML}{A4FF00}     
\definecolor{Emerald}{HTML}{50C878}
\usepackage{amsmath}
\usepackage[british]{babel}
\usepackage{graphicx}
\usepackage{dcolumn}
\usepackage{ragged2e}
\usepackage{bm}
\usepackage[version=4]{mhchem}
\usepackage{comment}
\usepackage{hepparticles}
\usepackage{placeins}
\usepackage{booktabs}
\usepackage{lipsum}
\usepackage{afterpage}
\usepackage{tikz}
\usepackage{ulem}
\usepackage{hyperref}
\hypersetup{
    colorlinks=true,
    linkcolor= VividBlue,
    filecolor=cyan,      
    urlcolor= VividBlue,
    citecolor= VividBlue,
    pdftitle={59Cu_PRC_coauthors},
    pdfpagemode=FullScreen,
    }
\usepackage{array}
\usepackage{soul}
\usepackage{orcidlink}

\newcommand{\redsout}[1]{%
    \bgroup
    \markoverwith{\textcolor{red}{\rule[0.5ex]{2pt}{1pt}}}%
    \ULon{#1}%
    \egroup
}
\newcommand{\stot}{$\sigma_{\rm{reac}}$}
\begin{document}
\flushbottom



\title{Detailed Study of the $^{59}$Cu(p,$\alpha)^{56}$Ni Reaction and Constraints on Its Astrophysical Reaction Rate}

\author{E. Lopez-Saavedra \orcidlink{0000-0001-9187-0404}}
\email{elopezsaavedra@anl.gov}
\affiliation{Physics\hspace{0.1cm}Division,\hspace{0.1cm}Argonne\hspace{0.1cm}National\hspace{0.1cm}Laboratory,\hspace{0.1cm}Lemont,\hspace{0.1cm}Illinois\hspace{0.1cm}60439,\hspace{0.1cm}USA}

\author{M. L. Avila \orcidlink{0009-0002-4051-9627}} 
\affiliation{Physics\hspace{0.1cm}Division,\hspace{0.1cm}Argonne\hspace{0.1cm}National\hspace{0.1cm}Laboratory,\hspace{0.1cm}Lemont,\hspace{0.1cm}Illinois\hspace{0.1cm}60439,\hspace{0.1cm}USA}

\author{W.-J. Ong \orcidlink{0000-0002-0945-8654}}
\affiliation{Nuclear\hspace{0.1cm}and\hspace{0.1cm}Chemical\hspace{0.1cm}Sciences\hspace{0.1cm}Division,\hspace{0.1cm}Lawrence\hspace{0.1cm}Livermore\hspace{0.1cm}National\hspace{0.1cm}Laboratory,\hspace{0.1cm}Livermore,\hspace{0.1cm}California\hspace{0.1cm}94550,\hspace{0.1cm}USA}

\author{P. Mohr \orcidlink{0000-0002-6695-9359}}
\affiliation{HUN-REN Institute for Nuclear Research (ATOMKI), H-4001 \hspace{0.1cm} Debrecen, Hungary}

\author{S. Ahn \orcidlink{0000-0001-8190-4914}}
\affiliation{Center\hspace{0.1cm}for\hspace{0.1cm}Exotic\hspace{0.1cm}Nuclear\hspace{0.1cm}Studies,\hspace{0.1cm}Institute\hspace{0.1cm}for\hspace{0.1cm}Basic\hspace{0.1cm}Science,\hspace{0.1cm}Daejeon\hspace{0.1cm}34126,\hspace{0.1cm}Republic\hspace{0.1cm}of\hspace{0.1cm}Korea}

\author{H. Arora \orcidlink{0000-0001-6183-658X}}
\affiliation{Center\hspace{0.1cm}for\hspace{0.1cm}Exotic\hspace{0.1cm}Nuclear\hspace{0.1cm}Studies,\hspace{0.1cm}Institute\hspace{0.1cm}for\hspace{0.1cm}Basic\hspace{0.1cm}Science,\hspace{0.1cm}Daejeon\hspace{0.1cm}34126,\hspace{0.1cm}Republic\hspace{0.1cm}of\hspace{0.1cm}Korea}

\author{L. Balliet \orcidlink{0000-0002-1081-1259}}
\affiliation{Facility\hspace{0.1cm}for\hspace{0.1cm}Rare\hspace{0.1cm}Isotope\hspace{0.1cm}Beams\hspace{0.1cm}(FRIB),\hspace{0.1cm}Michigan\hspace{0.1cm}State\hspace{0.1cm}University,\hspace{0.1cm}East\hspace{0.1cm}Lansing,\hspace{0.1cm}MI\hspace{0.1cm}48824,\hspace{0.1cm}USA}

\author{K. Bhatt \orcidlink{0000-0003-0100-8736}}
\affiliation{University of Notre Dame, Dept.\hspace{0.1cm} of\hspace{0.1cm} Physics\hspace{0.1cm} and \hspace{0.1cm} Astronomy, Notre \hspace{0.1cm}Dame, IN \hspace{0.1cm}46556, \hspace{0.1cm}USA}

\author{S.M. Cha \orcidlink{0009-0000-5988-5956}}
\affiliation{Center\hspace{0.1cm}for\hspace{0.1cm}Exotic\hspace{0.1cm}Nuclear\hspace{0.1cm}Studies,\hspace{0.1cm}Institute\hspace{0.1cm}for\hspace{0.1cm}Basic\hspace{0.1cm}Science,\hspace{0.1cm}Daejeon\hspace{0.1cm}34126,\hspace{0.1cm}Republic\hspace{0.1cm}of\hspace{0.1cm}Korea}

\author{K. A. Chipps \orcidlink{0000-0003-3050-1298}}
\affiliation{Oak\hspace{0.1cm}Ridge\hspace{0.1cm}National\hspace{0.1cm}Laboratory,\hspace{0.1cm}Tennessee,\hspace{0.1cm}USA}

\author{J. Dopfer \orcidlink{0009-0007-9235-4824}}
\affiliation{Facility\hspace{0.1cm}for\hspace{0.1cm}Rare\hspace{0.1cm}Isotope\hspace{0.1cm}Beams\hspace{0.1cm}(FRIB),\hspace{0.1cm}Michigan\hspace{0.1cm}State\hspace{0.1cm}University,\hspace{0.1cm}East\hspace{0.1cm}Lansing,\hspace{0.1cm}MI\hspace{0.1cm}48824,\hspace{0.1cm}USA}

\author{I. A. Tolstukhin \orcidlink{0000-0002-6631-7479}}
\affiliation{Physics\hspace{0.1cm}Division,\hspace{0.1cm}Argonne\hspace{0.1cm}National\hspace{0.1cm}Laboratory,\hspace{0.1cm}Lemont,\hspace{0.1cm}Illinois\hspace{0.1cm}60439,\hspace{0.1cm}USA}

\author{R. Jain \orcidlink{0000-0001-9859-1512}}
\affiliation{Nuclear\hspace{0.1cm}and\hspace{0.1cm}Chemical\hspace{0.1cm}Sciences\hspace{0.1cm}Division,\hspace{0.1cm}Lawrence\hspace{0.1cm}Livermore\hspace{0.1cm}National\hspace{0.1cm}Laboratory,\hspace{0.1cm}Livermore,\hspace{0.1cm}California\hspace{0.1cm}94550,\hspace{0.1cm}USA}

\author{M.J. Kim \orcidlink{0000-0002-4372-5592}}
\affiliation{Center\hspace{0.1cm}for\hspace{0.1cm}Exotic\hspace{0.1cm}Nuclear\hspace{0.1cm}Studies,\hspace{0.1cm}Institute\hspace{0.1cm}for\hspace{0.1cm}Basic\hspace{0.1cm}Science,\hspace{0.1cm}Daejeon\hspace{0.1cm}34126,\hspace{0.1cm}Republic\hspace{0.1cm}of\hspace{0.1cm}Korea}
\affiliation{Extreme\hspace{0.1cm}Rare\hspace{0.1cm}Isotope\hspace{0.1cm}Science,\hspace{0.1cm}Institute\hspace{0.1cm}for\hspace{0.1cm}Rare\hspace{0.1cm}Isotope\hspace{0.1cm}Science\hspace{0.1cm}(IRIS),\hspace{0.1cm}1\hspace{0.1cm}Gukjegwahak-ro,\hspace{0.1cm}Yuseong-gu,\hspace{0.1cm}Daejeon\hspace{0.1cm}34000,\hspace{0.1cm}Republic\hspace{0.1cm}of\hspace{0.1cm}Korea}

\author{K. Kolos \orcidlink{0000-0002-1726-4171}}
\affiliation{Nuclear\hspace{0.1cm}and\hspace{0.1cm}Chemical\hspace{0.1cm}Sciences\hspace{0.1cm}Division,\hspace{0.1cm}Lawrence\hspace{0.1cm}Livermore\hspace{0.1cm}National\hspace{0.1cm}Laboratory,\hspace{0.1cm}Livermore,\hspace{0.1cm}California\hspace{0.1cm}94550,\hspace{0.1cm}USA}

\author{F. Montes \orcidlink{0000-0001-9849-5555}}
\affiliation{Facility\hspace{0.1cm}for\hspace{0.1cm}Rare\hspace{0.1cm}Isotope\hspace{0.1cm}Beams\hspace{0.1cm}(FRIB),\hspace{0.1cm}Michigan\hspace{0.1cm}State\hspace{0.1cm}University,\hspace{0.1cm}East\hspace{0.1cm}Lansing,\hspace{0.1cm}MI\hspace{0.1cm}48824,\hspace{0.1cm}USA}

\author{D. Neto \orcidlink{0000-0002-5397-7048}}
\affiliation{Physics\hspace{0.1cm}Division,\hspace{0.1cm}Argonne\hspace{0.1cm}National\hspace{0.1cm}Laboratory,\hspace{0.1cm}Lemont,\hspace{0.1cm}Illinois\hspace{0.1cm}60439,\hspace{0.1cm}USA}
\affiliation{Department\hspace{0.1cm}of\hspace{0.1cm}Physics,\hspace{0.1cm}University\hspace{0.1cm}of\hspace{0.1cm}Illinois\hspace{0.1cm}Chicago,\hspace{0.1cm}845\hspace{0.1cm}W.\hspace{0.1cm}Taylor\hspace{0.1cm}St.,\hspace{0.1cm}Chicago,\hspace{0.1cm}IL\hspace{0.1cm}60607,\hspace{0.1cm}USA}

\author{S. D. Pain \orcidlink{0000-0003-3081-688X}}
\affiliation{Oak\hspace{0.1cm}Ridge\hspace{0.1cm}National\hspace{0.1cm}Laboratory,\hspace{0.1cm}Tennessee,\hspace{0.1cm}USA}

\author{J. Pereira \orcidlink{0000-0002-3934-0876}}
\affiliation{Facility\hspace{0.1cm}for\hspace{0.1cm}Rare\hspace{0.1cm}Isotope\hspace{0.1cm}Beams\hspace{0.1cm}(FRIB),\hspace{0.1cm}Michigan\hspace{0.1cm}State\hspace{0.1cm}University,\hspace{0.1cm}East\hspace{0.1cm}Lansing,\hspace{0.1cm}MI\hspace{0.1cm}48824,\hspace{0.1cm}USA}

\author{J. S. Randhawa \orcidlink{0000-0001-6860-3754}}
\affiliation{Department of Physics and Astronomy, Mississippi\hspace{0.1cm}State\hspace{0.1cm}University,\hspace{0.1cm}Mississippi\hspace{0.1cm}State,\hspace{0.1cm}MS\hspace{0.1cm}39762, USA}

\author{L. J. Sun \orcidlink{0000-0002-1619-7448}}
\affiliation{Facility\hspace{0.1cm}for\hspace{0.1cm}Rare\hspace{0.1cm}Isotope\hspace{0.1cm}Beams\hspace{0.1cm}(FRIB),\hspace{0.1cm}Michigan\hspace{0.1cm}State\hspace{0.1cm}University,\hspace{0.1cm}East\hspace{0.1cm}Lansing,\hspace{0.1cm}MI\hspace{0.1cm}48824,\hspace{0.1cm}USA}

\author{C. Ugalde \orcidlink{0000-0003-2721-6728}}
\affiliation{Department\hspace{0.1cm}of\hspace{0.1cm}Physics,\hspace{0.1cm}University\hspace{0.1cm}of\hspace{0.1cm}Illinois\hspace{0.1cm}Chicago,\hspace{0.1cm}845\hspace{0.1cm}W.\hspace{0.1cm}Taylor\hspace{0.1cm}St.,\hspace{0.1cm}Chicago,\hspace{0.1cm}IL\hspace{0.1cm}60607,\hspace{0.1cm}USA}

\author{L. Wagner \orcidlink{0009-0000-9954-9658}}
\affiliation{Facility\hspace{0.1cm}for\hspace{0.1cm}Rare\hspace{0.1cm}Isotope\hspace{0.1cm}Beams\hspace{0.1cm}(FRIB),\hspace{0.1cm}Michigan\hspace{0.1cm}State\hspace{0.1cm}University,\hspace{0.1cm}East\hspace{0.1cm}Lansing,\hspace{0.1cm}MI\hspace{0.1cm}48824,\hspace{0.1cm}USA}

\date{\today}

\begin{abstract}
The $^{59}$Cu$(p,\alpha)^{56}$Ni reaction plays an important role in explosive
astrophysical scenarios such as Type I X-ray bursts and the $\nu p$-process in
neutrino-driven winds following a core-collapse supernova, where it regulates
the flow of nucleosynthesis through the NiCu cycle and the synthesis of heavier
nuclei. We present a direct measurement of the
$^{59}\mathrm{Cu}(p,\alpha)^{56}\mathrm{Ni}$ excitation function from
2.43--5.88~MeV in the center-of-mass frame, performed in inverse kinematics
with the high-efficiency MUSIC active-target detector at FRIB. The angle- and
energy-integrated cross sections extend direct measurements to lower energies
than previously reported and remove the angular-integration model dependence of
earlier work. To extrapolate the rate to astrophysical energies, we constrain
the statistical-model description through a systematic optimization of the
DEM-3 $\alpha$-optical model potential geometry, and quantify the
model-selection uncertainty with a Bayesian model averaging analysis over 96
TALYS combinations. The resulting stellar rate carries a temperature-dependent
uncertainty factor of 1.26--1.63 over $T_9 = 0.2$--10 and is systematically
lower than the REACLIB evaluation, remaining below the competing $(p,\gamma)$
rate for $T_9 \lesssim 3.94$. These results substantially weaken the inferred NiCu
cycle strength and establish the $^{59}$Cu$(p,\gamma)^{60}$Zn rate as the
dominant remaining uncertainty.
\end{abstract}
\maketitle

\section{Introduction}
Proton-rich nucleosynthesis takes place in explosive astrophysical environments characterized by extreme temperatures and proton-rich compositions. These conditions are found in a variety of scenarios, such as Type~I X-ray bursts (XRBs) on accreting neutron stars \cite{WallaceWoosley1981,Parikh_2008} and proton-rich, neutrino-driven winds following core-collapse supernovae (CCSNe)~\cite{Frohlich2012_vpprocess,Arcones2012NuP}. While the overall astrophysical environment, such as temperature, density, and timescale, sets the stage for nucleosynthesis, it is the nuclear physics that dictates the detailed path of element formation \cite{Rauscher2012}. In both environments, the reaction path is highly sensitive to the competition between proton-capture and particle-emission reactions on unstable nuclei, making individual reaction rates critical for shaping the final abundance pattern.

XRBs are thermonuclear flashes occurring in the envelope of a neutron star accreting material from a companion star and represent the most commonly observed stellar explosions \cite{Parikh_2008}. These events are detected by space-based X-ray observatories through their characteristic burst light curves, which record the burst intensity as a function of time \cite{Lewin1993}. When combined with stellar models, these observations provide insights into neutron star properties, which represent the densest known form of visible matter in the Universe \cite{Woosley1976,TimingNeutronStars2021,PARIKH2013225,Degenaar2018}. The thermonuclear runaway in XRBs is powered by the $rp$-process and the $\alpha p$-process, involving proton- and $\alpha$-induced reactions on stable and radioactive nuclei formed from accreted hydrogen and helium. Numerous sensitivity studies have investigated the impact of nuclear reaction-rate uncertainties on XRB observables, particularly burst light curves \cite{Cyburt_2016,Meisel_2019}. These studies highlight that uncertainties in these reaction rates hinder quantitative comparisons between XRB models and observations.

The $\nu p$-process has been proposed as a potential site for the synthesis of elements beyond iron. It operates in proton-rich supernova ejecta, where antineutrino captures on free protons generate a small but crucial neutron density \cite{Frolich_nup}. These neutrons enable $(n,p)$ reactions that bypass $\beta^+$ waiting points, allowing the reaction flow to proceed to heavier nuclei. Subsequent studies have shown that the $\nu p$-process can efficiently produce nuclei with mass numbers $A > 64$ at temperatures below approximately $3$~GK \cite{Arcones2012NuP}.

Among the most influential reactions in both, the $rp$- and $\nu p$-processes, is the $^{59}$Cu$(p,\alpha)^{56}$Ni reaction \cite{Cyburt_2016,Arcones2012NuP}. Although the astrophysical conditions in these two scenarios differ substantially, both environments encounter a common bottleneck associated with the so-called NiCu cycle. At temperatures below approximately $3$~GK, the $^{59}$Cu$(p,\gamma)^{60}$Zn reaction dominates, allowing the reaction flow to proceed toward heavier nuclei. At higher temperatures, however, the $^{59}$Cu$(p,\alpha)^{56}$Ni reaction becomes dominant, confining the reaction flow to a closed NiCu cycle and inhibiting the synthesis of heavier elements. Consequently, the relative strengths of the $^{59}$Cu$(p,\gamma)^{60}$Zn and $^{59}$Cu$(p,\alpha)^{56}$Ni reactions play a decisive role in determining whether nucleosynthesis can proceed beyond the iron group in these proton-rich environments.

Despite its astrophysical importance, experimental data for the $^{59}$Cu$(p,\alpha)^{56}$Ni reaction remain scarce at low energies. The first direct measurement was performed by Randhawa \textit{et al.}~\cite{Jaspreet59Cu}, who reported a single data point at $E_{\mathrm{cm}} = 6.0 \pm 0.3~\mathrm{MeV}$. The measured cross section was a factor of $\sim$1.6–4 lower than statistical-model predictions; however, no reaction rate was extracted. More recently, Bhathi \textit{et al.}~\cite{Bhathi25} reported two additional data points between 4 and 5~MeV, indicating a reaction rate approximately a factor of two lower than that adopted in REACLIB. This previous measurement detected $\alpha$ particles over a limited angular range and found that the extracted angle-integrated cross sections were strongly sensitive to the assumed angular distribution. Other works \cite{Kim_2022,Lotay} have attempted to constrain this reaction rate using spectroscopic information on states in $^{60}\mathrm{Zn}$ and shell-model calculations. Ref.~\cite{Kim_2022} used available data from several transfer-reaction measurements, while Ref.~\cite{Lotay} used information from their recently measured $^{59}$Cu$(d,n)^{60}$Zn reaction. However, the resulting constraints suffer from significant uncertainties, primarily due to the lack of experimental data on the corresponding $\alpha$-particle partial widths. A series of experiments is planned and ongoing to study the NiCu cycle; a comprehensive overview was provided recently in Ref.~\cite{Sun_2025}.

\par
In this work, we present a new study of the $^{59}\mathrm{Cu}(p,\alpha)^{56}\mathrm{Ni}$ reaction, directly measuring angle- and energy-integrated cross sections at lower energies than previously explored. This approach removes the model dependence in the determination of the total cross sections and provides the strongest constraints to date on the astrophysical reaction rate, significantly reducing the associated uncertainties.
The paper is organized as follows: the experiment and results are
presented in Secs.~\ref{sec:experiment} and~\ref{sec:results}, the
theoretical analysis and reaction-rate determination in
Sec.~\ref{sec:theory}, a comparison with previous rates and astrophysical
implications in Sec.~\ref{sec:comparison}, and the conclusions in
Sec.~\ref{sec:conclusions}.

\section{Experiment}\label{sec:experiment}
The experiment was carried out in inverse kinematics at the Facility for Rare Isotope Beams (FRIB), located at Michigan State University (MSU). For the measurement, the Multi-Sampling Ionization Chamber (MUSIC) detector~\cite{Carnelli2015MUSIC}, which was developed at the Argonne Tandem Linac Accelerator (ATLAS) facility, was used. The $^{59}$Cu beam was produced in flight by fragmentation of a 240 MeV/u $^{64}$Zn primary beam, then stopped and reaccelerated by the reaccelerator (ReA) to an energy of 8.418 MeV/u. The MUSIC detector was installed on the ReA6 beamline. The $^{59}$Cu beam incident on the MUSIC detector had an average purity of approximately 94\% and an intensity of $\sim9\times10^{3}$ particles per second. The only beam contaminant was $^{59}$Ni, contributing approximately $6\%$ of the total beam. 
The MUSIC detector was filled with 440 Torr of methane gas (CH$_4$), allowing the measurement of an excitation function over a wide energy range. At this pressure, the beam was completely stopped within the detector. 

\begin{figure}
        \centering
        \includegraphics[width=0.9\linewidth]{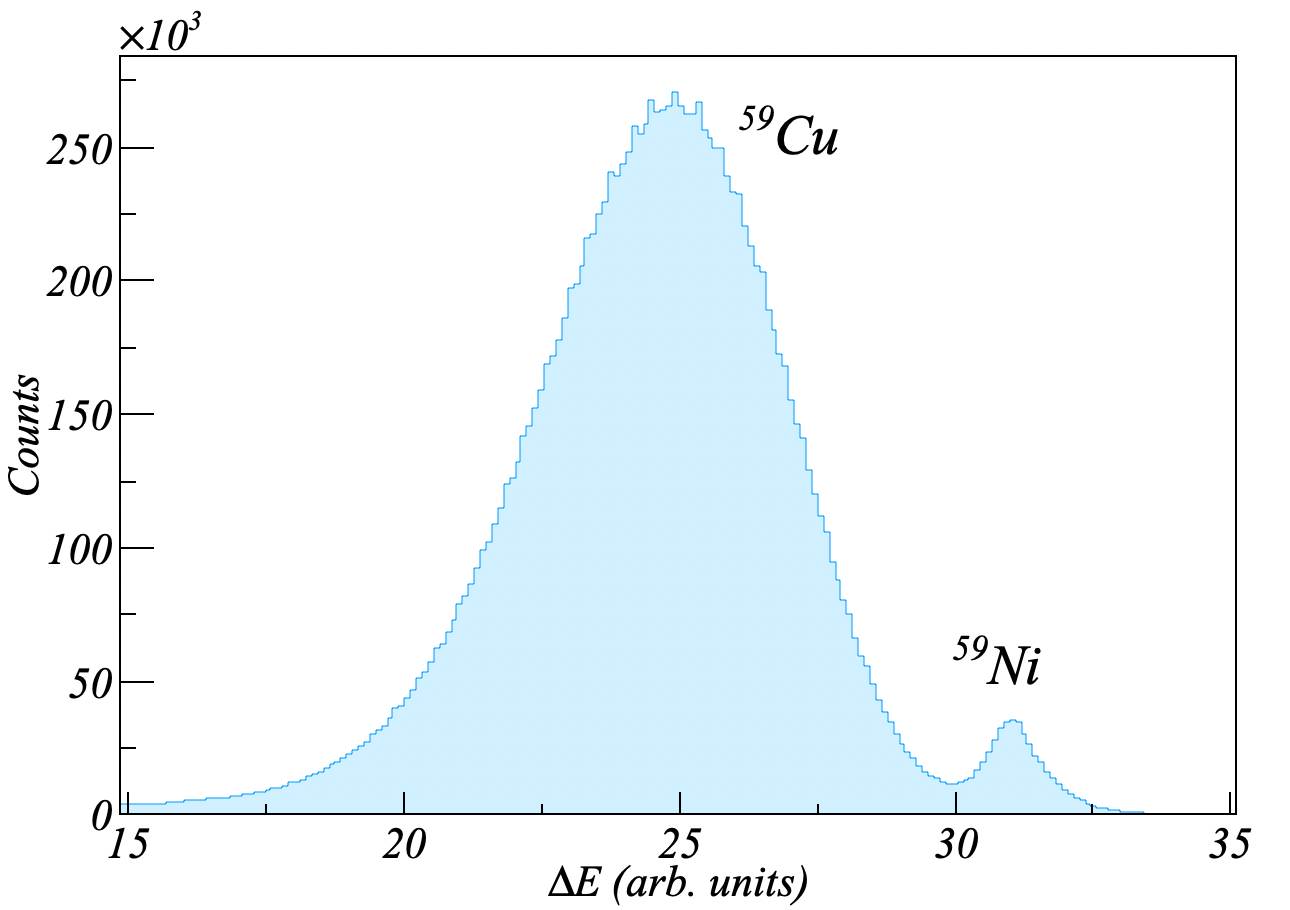}
      \caption{Beam profile on Strip~12 (right) of the MUSIC anode.
The \textsuperscript{56}Ni beam contaminant is cleanly separated from the primary \textsuperscript{59}Cu component.}
        \label{fig:beamprof}
    \end{figure}

The MUSIC detector is a highly efficient active target system (close to 100\% efficiency) with an anode segmented into 18 strips (0–17). The middle strips, 1–16, are further subdivided into right and left sections. A complete description of the detector and a detailed discussion of its operating principles are given in Refs.~\cite{Carnelli2015MUSIC,AvilaNIM}. MUSIC provides excellent particle identification by measuring the energy loss of ions across multiple anode segments, allowing separation according to atomic number ($Z$) and distinguishing different reactions from beam contaminants. Fig.~\ref{fig:beamprof} demonstrates that the $^{59}$Ni component is fully resolved from the primary $^{59}$Cu beam in the beam-profile spectrum when using anode strip 12.

\begin{figure*}[!t]
    \centering
    \begin{tabular}{cc}
        \includegraphics[width=0.46\textwidth]{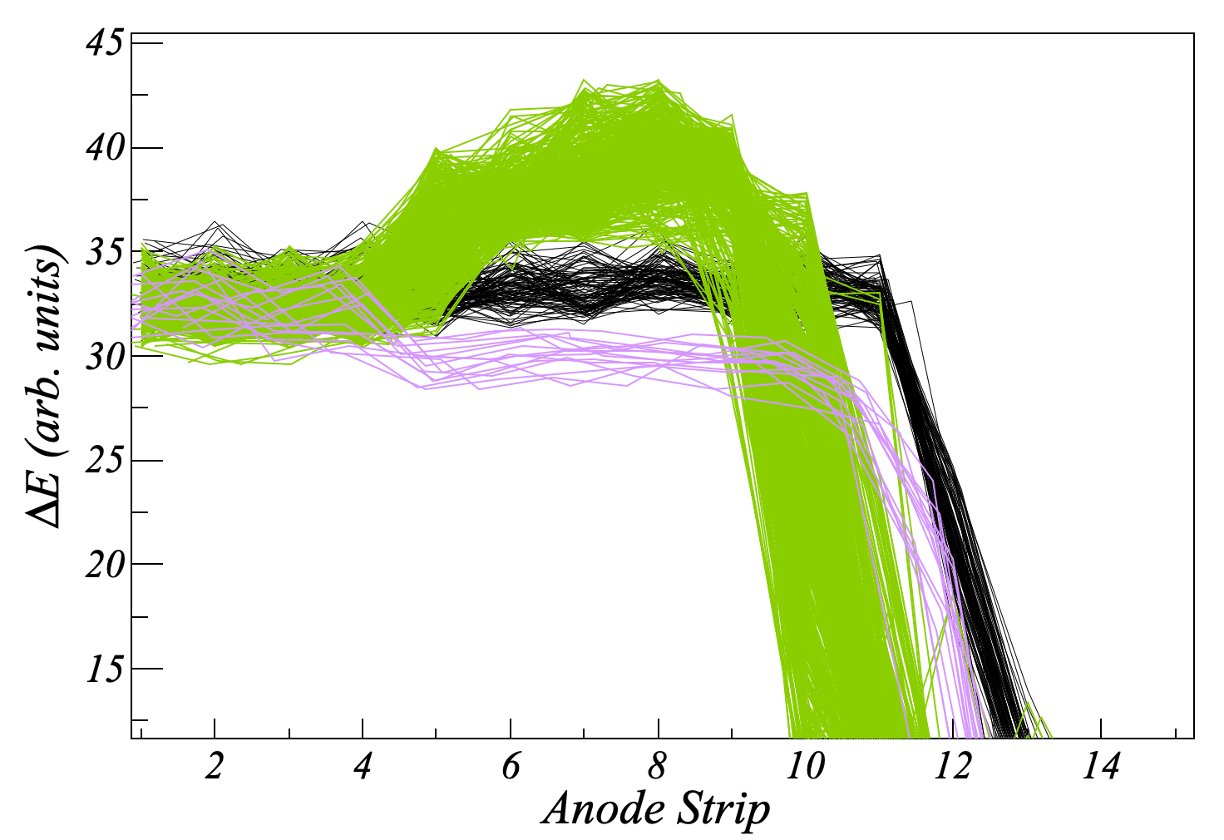} &
        \includegraphics[width=0.4\textwidth]{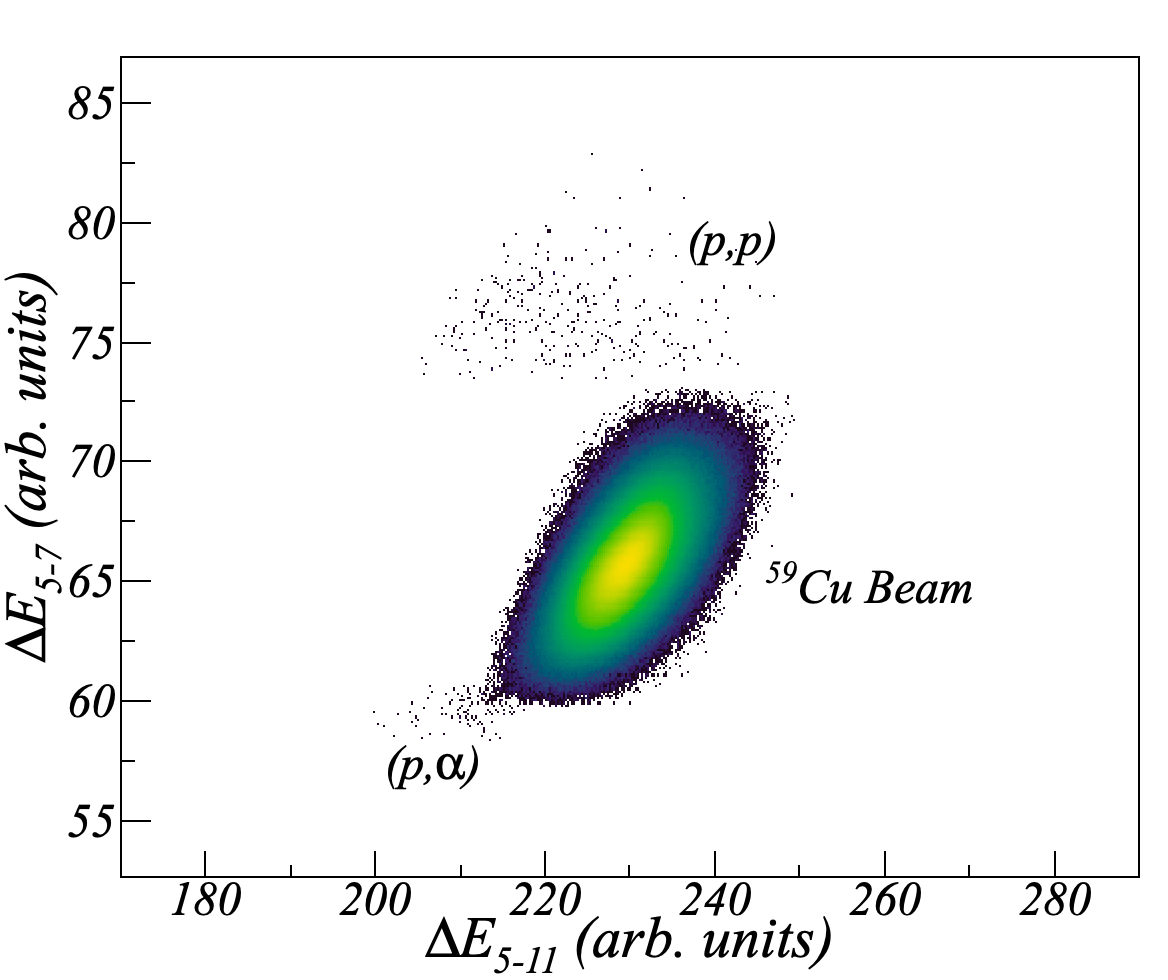} \\
    \end{tabular}
    \caption{(left) Energy loss as a function of strip number "traces" of $(p,\alpha)$ events occurring in strip 5 (purple line), the unreacted $^{59}$Cu beam (black line), and the $(p,p)$ and $(p,p')$ events (green line). (right) $\Delta E$--$\Delta E$ spectrum of events occurring in strip 5.}
    \label{fig:pa_identification}
\end{figure*}
To identify $(p,\alpha)$ events, MUSIC measures differences in energy loss across multiple anode strips after a reaction, producing distinct “traces” that distinguish the heavy recoil $^{56}$Ni from the unreacted $^{59}$Cu beam due to its lower energy deposition. At these energies, other reaction channels such as $(p,p)$, $(p,p')$, and $(p,\gamma)$ are energetically allowed. However, the recoiling $^{59}$Cu and $^{60}$Zn nuclei have different energy losses and do not interfere with the identification of $(p,\alpha)$ events. Fig.~\ref{fig:pa_identification} (left) shows the traces of $(p,\alpha)$ events occurring in strip 5 (blue line), the unreacted $^{59}$Cu beam (black line), and the $(p,p)$ and $(p,p')$ events (green line). For simplicity, the energy loss measured across strips 2–11 is normalized to that of the $^{59}$Cu beam in strip 1.

As a parallel event-identification approach, the segmented anode of MUSIC allows the energy deposited in consecutive strips downstream of the reaction point to be arranged in a $\Delta E$--$\Delta E$ configuration. This representation allows the distinct energy-loss characteristics of the reaction products and enables independent identification of the different reaction channels on an event-by-event basis. Fig.~\ref{fig:pa_identification} (right) shows a representative $\Delta E$--$\Delta E$ spectrum, in which the $(p,\alpha)$ events are clearly separated from the unreacted beam as well as from elastic and inelastic scattering events, for events occurring in strip~5.

 \section{Experimental Results}\label{sec:results}

The excitation function was measured at effective center-of-mass energies between 2.43 and 5.88 MeV. The measured effective energies, energy binning, and cross sections with statistical and systematic uncertainties corresponding to anode strips 2 to 9 of the MUSIC detector are shown in Table~\ref{tab:xsec_results}. 
Effective center-of-mass energies, $E_{\text{c.m.,eff}}$, were calculated by first deducing the center-of-mass energies from the measured beam energy losses in the MUSIC detector and then applying corrections for thick-target yield effects using the following equation from Ref.~\cite{Cauldrons}: 
\begin{equation}
\begin{aligned}
E_{\mathrm{eff}} &= E_{0} - \Delta E + \Delta E \left[
-\frac{\sigma_{2}}{\sigma_{1}-\sigma_{2}}
+ \sqrt{\frac{\sigma_{1}^{2}+\sigma_{2}^{2}}{2(\sigma_{1}-\sigma_{2})^{2}}}
\right]
\end{aligned}
\label{eq:Eeff}
\end{equation}

where $\sigma_{1}$ and $\sigma_{2}$ are the cross sections at the anode strip entrance and exit, $E_0$ is the energy at the beginning of the strip, and $\Delta E$ is the energy loss in the strip. Eq.~\ref{eq:Eeff} provides a good approximation for $\sigma_{1}/\sigma_{2} \lesssim 10$. In our measurements, this ratio is always $\sigma_{1}/\sigma_{2} \lesssim 5$. The cross-section weighting was evaluated using the statistical model calculations with the Demetriou–3 (DEM–3) optical model potential, as described in Sec.~\ref{sec:TALYS}.

The center-of-mass energy uncertainty originates from the beam energy spread (0.5\% FWHM) and the uncertainty in the beam energy loss across the detector strips ($\sim$5\%).
The latter was estimated by using different energy-loss tables (such as ATIMA~\cite{ATIMA} and Ziegler~\cite{Ziegler}) in LISE++~\cite{lise}, and adjusting them to match the beam's Bragg curve. It was observed that the Ziegler tables reproduced the Bragg curve, whereas for ATIMA the calculated energy loss had to be reduced by approximately 5\%. Therefore, an uncertainty of 5\% was adopted for the energy-loss calculation, resulting in changes to the center-of-mass energy ranging from 0.6\% for the first strip to about 2\% for the last strip shown in Table~\ref{tab:xsec_results}.
The estimated total energy loss of the beam in each anode strip defines the center-of-mass energy binning, $\Delta E_{c.m}$. 

The systematic uncertainty of the cross sections originates primarily from the analysis methods and the different criteria used to select the events of interest. Systematic uncertainties varied by strip, averaging below 17\% for this experiment, except for strips 2 and 6, which were affected by noise and exhibited uncertainties of approximately 33\%. The larger systematic uncertainties observed in this experiment, relative to previous MUSIC measurements, are attributed to the higher instantaneous beam rate, which led to degraded detector resolution due to space-charge effects.

\begin{table}[!htb]
\centering
\caption{Measured effective energies, energy bins, and cross sections with statistical and systematic uncertainties.}
\label{tab:xsec_results}
\begin{tabular}{c c c c c}
\toprule
$E_{\mathrm{cm}}^{eff}$\hspace{-0.07cm} (MeV)&$\Delta E_{cm}$\hspace{-0.07cm} (MeV)& $\sigma$\hspace{-0.07cm} (mb)&$\Delta\sigma_{\text{stat}}$\hspace{-0.07cm} (mb)& $\Delta\sigma_{\text{sys}}$\hspace{-0.07cm} (mb) \\
\midrule
5.88 (3) &$^{+0.18}_{-0.26}$ & $7.024$     & $0.220$   & $2.355$ \\
5.44 (3)& $^{+0.18}_{-0.27}$ & $3.412$     & $0.154$   & $0.612$ \\
4.98 (4)& $^{+0.19}_{-0.28}$ & $1.861$   & $0.113$   & $0.300$ \\
4.49 (4) & $^{+0.21}_{-0.28}$ & $0.619$     & $0.065$  & $0.092$ \\
4.02 (4) & $^{+0.19}_{-0.31}$ & $0.248$     & $0.042$  & $0.083$ \\
3.54 (4) & $^{+0.17}_{-0.37}$ & $0.082$ & $0.024$  & $0.014$ \\
2.98 (5) & $^{+0.19}_{-0.37}$ & $\llap{\tiny\textless}0.015$   & $-$ & $-$ \\
2.43 (5) & $^{+0.18}_{-0.42}$ & $\llap{\tiny\textless}0.011$   & $-$ & $-$ \\
\bottomrule
\end{tabular}
\end{table}

For the strips that have a higher number of counts, Poisson statistics were used to estimate the statistical uncertainties of the cross sections. For strips with very low statistics, specifically strips~8 and~9, confidence intervals were determined using the Feldman--Cousins unified approach to classical confidence belts~\cite{FeldmanCousins1998}.
This construction avoids the ambiguity of deciding \textit{a priori} between upper limits and two-sided intervals and guarantees correct frequentist coverage even in the low-count regime. The method is based on the ordering principle defined by the likelihood ratio:
\begin{equation}
R(n;\mu) = \frac{P(n \mid \mu)}{P(n \mid \mu_{\text{best}})} 
\end{equation}
where $P(n \mid \mu)$ is the Poisson probability of observing $n$ events given signal mean $\mu$ and known background $b$, and $\mu_{\text{best}} = \max(0,\, n-b)$ is the signal strength that maximizes the likelihood subject to physical constraints. Values of $n$ are included in the acceptance interval in decreasing order of $R$ until the desired confidence level is reached. In this way, Feldman and Cousins constructed tables of confidence intervals such as Tables~VI and VII in Ref.~\cite{FeldmanCousins1998}. Specifically, for strip~8 with $n_{0}=2$ events and strip~9 with $n_{0}=1$ event, and assuming a known background of $b=5$, the 95$\%$~C.L.\ intervals for the signal mean are $[0.00,\,2.49]$ and $[0.00,\,1.88]$ events, respectively. 
These intervals reduce to 95\% C.L. upper limits when the event statistics are extremely low, while still preserving the correct frequentist coverage.

\begin{figure}[!ht]
    \centering
    \includegraphics[width=0.9\linewidth]{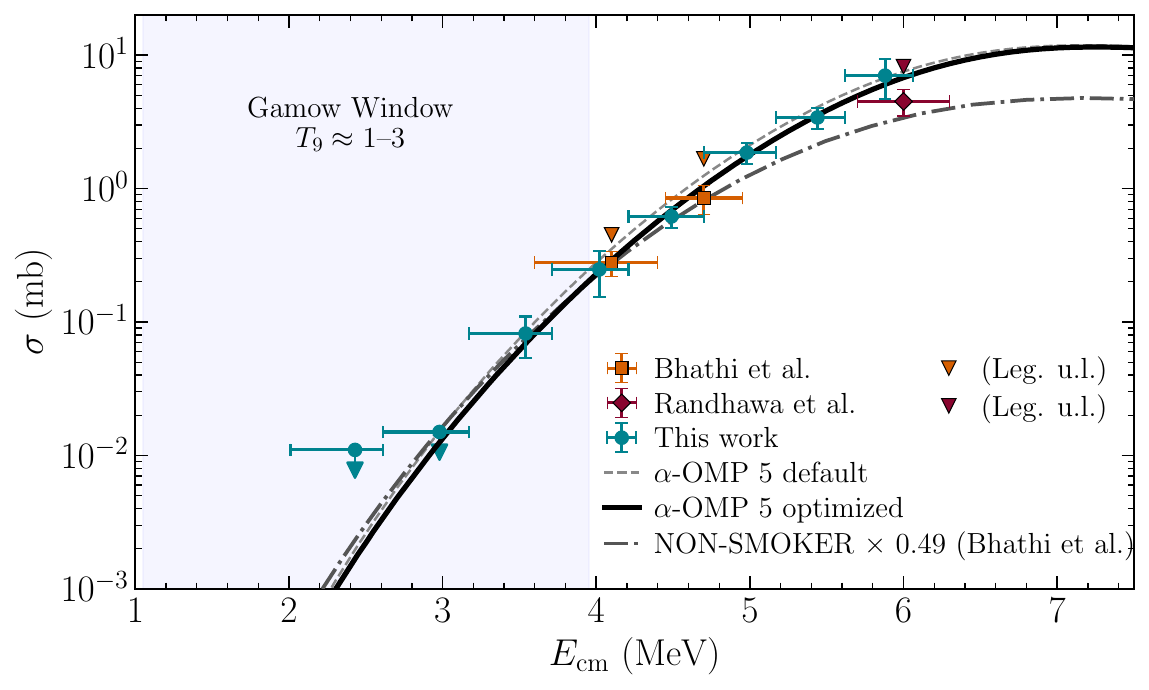}
   \caption{Measured $^{59}\mathrm{Cu}(p,\alpha)^{56}\mathrm{Ni}$ cross sections from this work (teal circles), Randhawa~\textit{et al.}~\cite{Jaspreet59Cu} (maroon diamond), and Bhathi~\textit{et al.}~\cite{Bhathi25} (orange squares: recommended values; triangles: Legendre upper limits), compared with the NON-SMOKER$\,\times\,0.49$ adopted by Bhathi \textit{et al.} (dark gray dash-dotted), and the DEM-3 ($\alpha$-OMP 5)~\cite{DEMETRIOU2002253} with optimized (black) and unmodified (gray dashed) geometry. Shaded region represents the Gamow window for temperatures $T_9\simeq1$--3}
    \label{fig:xsec_comparison}
\end{figure}

Fig.~\ref{fig:xsec_comparison} compares the measured cross sections with the previous direct measurements of Randhawa \textit{et al.}~\cite{Jaspreet59Cu} and Bhathi \textit{et al.}~\cite{Bhathi25}, together with three statistical-model predictions. The first is the NON-SMOKER prediction scaled by a factor of 0.49, following the normalization adopted in Ref.~\cite{Bhathi25}. The other two are
TALYS~\cite{talys2} calculations using the Demetriou and Goriely dispersive (DEM-3) $\alpha$-optical model potential~\cite{DEMETRIOU2002253}, shown with its default geometry and with the optimized geometry (see Sec.~\ref{sec:geom_opt});
the latter reproduces our data without any scaling factor and provides the best agreement with the present measurement. A detailed discussion of the statistical-model calculations is given in Secs.~\ref{sec:TALYS} and~\ref{sec:geom_opt}. \par
The measurements of Refs.~\cite{Jaspreet59Cu,Bhathi25} used the same experimental setup, in which the recoiling $\alpha$-particles were detected in an angular range of $\sim$18°-- 40° in the laboratory frame with a $\Delta E$–$E$ silicon detector telescope. At the energy closest to the Randhawa \textit{et al.} data point ($E_{\mathrm{cm}} \approx 6.0\pm0.3$~MeV), our measurement at $E_{\mathrm{cm}} = 5.88$~MeV yields a cross section roughly a factor of 1.5 higher, yet still consistent within the quoted uncertainties.
The measurement of Ref.~\cite{Bhathi25} reported two alternative extractions of the two angle-integrated cross section derived from the same experimental data set, depending on the assumed form of the angular distribution used in the integration procedure, namely, a TALYS-based angular distribution and a Legendre-polynomial fit to the measured angular data. Both extractions are shown in the present work to illustrate the resulting method-dependent spread associated with the angle-integration procedure. As shown in Fig.~\ref{fig:xsec_comparison}, the present result is consistent with the angle-integrated cross section obtained in Ref.~\cite{Bhathi25} when the TALYS-based angular distribution is used in the integration, while a larger deviation is observed for the extraction based on the Legendre-polynomial fit. The opposite is found for the one data point of Ref.~\cite{Jaspreet59Cu} at 6.0 MeV. Our measurement with the MUSIC detector provides total, energy- and angle-integrated cross sections and does not suffer from the systematic uncertainties of previous measurements.

\section{Theoretical Analysis}\label{sec:theory}

In this section, Hauser–Feshbach statistical-model calculations performed with the TALYS 2.0 code are used to extrapolate the experimental cross sections to lower energies and to derive reaction rates at astrophysical relevant temperatures. Since the present measurements probe only reactions proceeding from the ground state of the target nucleus $^{59}$Cu, the statistical-model calculations are also employed to evaluate the stellar enhancement factor, which accounts for contributions from thermally populated excited states in a stellar environment and allows the laboratory reaction rate to be converted into a stellar reaction rate.

\subsection{Statistical model calculations}\label{sec:TALYS}
The Hauser–Feshbach formalism~\cite{Hauser1952} provides a statistical description of compound-nucleus reactions, in which the reaction cross section is determined by the competition among the available decay channels, expressed in terms of transmission coefficients. In a simplified notation, the $(p,\alpha)$ cross section scales as
\begin{equation}
   \sigma_{(p,\alpha)} \propto \frac{T_{p0}\,T_{\alpha}}{T_p + T_\gamma + T_\alpha}
       \approx \frac{T_{p0}\,T_{\alpha 0}}{T_{p}} 
       \label{eq:trans}
\end{equation}
where the approximation holds for the energies under study. 
The $T_i$ are the transmissions to the open channels, summed over all final states $j$ in the residual nuclei, $T_i = \sum_j T_{i,j}$. The quantity $T_{p0}$ is the transmission in the entrance channel, with the target $^{59}$Cu in its ground state. The $\alpha$-particle transmission is well approximated by $T_\alpha \simeq T_{\alpha 0}$, as it is dominated by the transition to the $^{56}$Ni ground state; transitions to excited states are suppressed due to the lower $\alpha$-particle energy required to populate excited states in the doubly magic $^{56}$Ni nucleus. Furthermore, $T_p \gg T_\alpha$ results from the higher Coulomb barrier in the $^{59}$Ni + $\alpha$ channel than in the $^{59}$Cu + p channel.

We use the above simplistic Eq.~(\ref{eq:trans}) in the following discussion because it allows a simple and desciptive interpretation. The full calculations in the TALYS code \cite{talys2} include in addition the conservation of spin and parity and a correlation of the incoming and outgoing wave functions by a so-called width fluctuation factor (see e.g.\ \cite{Moldauer_PRC1976}).

Eq.~(\ref{eq:trans}) indicates that the calculated $(p,\alpha)$ cross section depends on the proton optical model potential (pOMP) through $T_p$ and $T_{p0}$, on the $\alpha$-OMP through $T_{\alpha 0}$, and on the level density model (ld) in $^{59}$Cu through $T_p$. For further discussion of the role of the transmissions $T_i$, see Ref.~\cite{Mohr_EPJA2025_aomp}. The influence of the pOMP remains minor because variations in the proton transmission affect $T_{p0}$ and $T_p$ in a similar way, resulting in an almost unchanged ratio $T_{p0}/T_p$ in Eq.~(\ref{eq:trans}). Consequently, to identify the optimal settings for extrapolating the cross section, one has to focus on the $\alpha$-OMP and the ld.

\begin{figure}[!ht]
    \centering
    \includegraphics[width=0.9\linewidth]{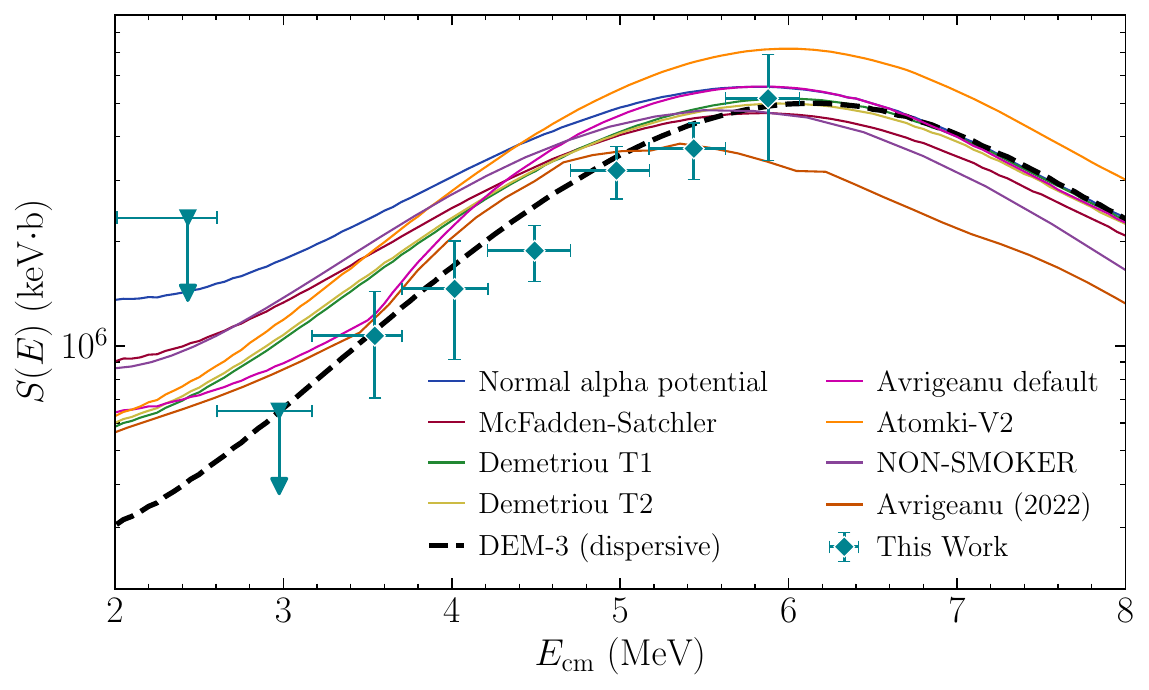}
 \caption{Measured $^{59}\mathrm{Cu}(p,\alpha)^{56}\mathrm{Ni}$ S-factors from the present work (red diamonds), compared with TALYS calculations using different $\alpha$-OMP, NON-SMOKER and the new pOMP derived for this reaction by Avrigeanu (2022)~\cite{Avrigeanu2022}.}
    \label{fig:TALYS_AOMP}
\end{figure}

To identify the $\alpha$-OMPs that best reproduce the experimental values, TALYS calculations ~\cite{talys2} were performed using different $\alpha$-OMPs, and the results are shown in Fig.~\ref{fig:TALYS_AOMP}. The Nolte~\cite{Nolte1987_aomp} and Avrigeanu (variant)~\cite{Avrigeanu1994_aomp_variant} $\alpha$-OMPs were omitted, as these potentials were constrained at higher energies and are therefore not appropriate for astrophysical applications. All $\alpha$-OMPs tend to overestimate the experimental $(p,\alpha)$ cross sections (as well as the S-factors), as previously noted in Ref.~\cite{Jaspreet59Cu}. Among them, the DEM-3~\cite{DEMETRIOU2002253} dispersive $\alpha$-OMP provides the best agreement with the data. A consistent performance of this potential was also reported by Gyürky {\it{et al.}}\ in their study of the $^{64}$Zn($p,\alpha$)$^{61}$Cu reaction~\cite{Gy}, where TALYS calculations employing the same Demetriou potential showed good agreement with experimental results. The upper limit at the lowest energy is too high to impose meaningful constraints on the $\alpha$-OMP. In contrast, the upper limit near 3~MeV requires a relatively steep energy dependence, excluding most of the $\alpha$-OMPs considered and clearly favoring the Demetriou and Goriely dispersive $\alpha$-OMP as well as the Atomki-V2 potential~\cite{Mohr_PRL2020_AtomkiV2,Mohr_ADNDT2021_AtomkiV2}.

An alternative approach by Avrigeanu~\cite{Avrigeanu2022}, which considered pOMP anomalies at sub-Coulomb energies and modified the pOMP to reproduce the data of Ref.~\cite{Jaspreet59Cu}, leads to a somewhat flatter energy dependence at higher energies. This flatter energy dependence arises from the reduced energy sensitivity associated with the modified proton imaginary potential, as illustrated by the excitation functions in Ref.~\cite{Avrigeanu2022}, and contrasts with the steeper behavior obtained when variations are driven primarily by the $\alpha$-OMP.

\subsection{Bayesian Model Averaging of Statistical-Model Calculations}

Fig.~\ref{fig:TALYS_AOMP} already hints that DEM-3 provides the best
reproduction of the measured cross sections among the $\alpha$-OMP
parametrizations considered. To rigorously assess which model combination is best supported by the data and propagate model-selection uncertainty into the stellar rate, we performed a Bayesian model averaging (BMA) analysis~\cite{Hoeting1999} over all eight
$\alpha$-OMP parametrizations available in TALYS-2.0, combined with systematic
variations of the level-density model (ld = 1--6) and the number of discrete
states included in the target nucleus $^{59}$Cu (msl = 5 and 30), yielding 96
model combinations in total. BMA has previously been applied to nuclear
reaction data evaluation with TALYS~\cite{Alhassan2024}, and related Bayesian
uncertainty-quantification methods have been used to propagate nuclear-model
uncertainties into astrophysical reaction rates~\cite{Chalil2024}.

\par

In TALYS, the parameters  msl and ld control how the compound-nucleus transmission coefficients are constructed. As stated above, msl indicates the amount of discrete levels included in the target nucleus. Above the energy of the highest included level, the level density model is used to generate a quasi-continuum of states. Thus, decreasing the amount of included discrete states (e.g., from 30 to 5) makes the proton transmission coefficient $T_{p}$ more sensitive to the assumed level density, because the level-density formula is ``called"  at lower excitation energies. Since the $(p,\alpha)$ cross section depends strongly on $T_{p}$ as seen in Eq.~(\ref{eq:trans}), variations in ld and msl primarily impact the population of excited states in $^{59}$Cu, which dominates the denominator of that equation via elastic and inelastic proton channels. As a result, these parameters govern the largest model-dependent variations in the calculated $(p,\alpha)$ cross sections. As will be shown below, the choice of the lower msl = 5 will result in a better reproduction of the experimental data. When the higher msl = 30 is used, one implicitly relies on that the experimental level scheme of the unstable $^{59}$Cu nucleus is complete up to the known $30^{\rm{th}}$ excited state which is located at an excitation energy of about 3.5 MeV; but it is experimentally very difficult to ensure the completeness of a level scheme because weak or overlapping levels may easily be overlooked.

Variations in the level densities of the other nuclei involved ($^{56}$Ni, $^{60}$Zn) have only minor influence, as $T_{\alpha}$ is dominated by the ground-state transmission and the $\gamma$-decay channel remains weak, so changes in $T_{\gamma}$ have negligible effect on the $(p,\alpha)$ cross section. In the specific case of $^{60}$Zn, Soltesz \textit{et al.}~\cite{Soltesz} studied its level density via the $^{58}$Ni($^{3}$He,n)$^{60}$Zn reaction and reported an unusual level-density shape at excitation energies of about 5–6 MeV. This feature raises concerns about the applicability of the statistical model for calculating reaction rates at low temperatures ($T_9 \lesssim 1$).

In the BMA analysis, each model combination $i$ was assigned a marginal
likelihood weight
\begin{equation}
\label{eq:bma_weight}
  w_i \;\propto\; \frac{\exp(-\chi^2_i/2)}{\sqrt{A_i}},
\end{equation}
where $\chi^2_i$ is the minimum chi-squared of model combination $i$,
evaluated at its best-fit scaling factor $k$, and $A_i$ is the Fisher
information for that scaling factor,
\begin{equation}
  A_i \;=\; \sum_j \frac{f_i(E_j)^2}{\sigma_j^2},
\end{equation}
where $f_i(E_j)$ is the unscaled prediction of model $i$ at energy $E_j$
and $\sigma_j$ is the uncertainty on the measured cross section. The
weight in Eq.~(\ref{eq:bma_weight}) is the exact marginal likelihood for a model linear in
its amplitude, $\sigma_{\mathrm{theory}} = k\,f_i(E)$~\cite{Sivia2006}:
since $\chi^2(k)$ is quadratic in $k$, the Gaussian likelihood can be
integrated analytically over $k$, assuming no prior preference for any particular value of the scaling factor,
\begin{equation}
  \int_{-\infty}^{\infty}
  \exp\!\left(-\frac{A_i\,(k-k_{\mathrm{best}})^2}{2}\right) dk
  \;=\; \sqrt{\frac{2\pi}{A_i}},
\end{equation}
and the resulting $1/\sqrt{A_i}$ factor is the Occam penalty for the
fitted normalization~\cite{Trotta2008}: a model that reproduces the data only over a narrow range of $k$ (large $A_i$) has effectively consumed a fitting parameter and is downweighted relative to one that fits without strong dependence on $k$. The same equal weighting of $k$ is used for all $96$ combinations, so the (formally divergent) normalization cancels in the weights $w_i$, which therefore depend only on the relative evidence between models. Each of the $96$ model combinations was assigned equal probability beforehand, so that the posterior weights are determined entirely by the marginal likelihood.
DEM-3 emerges as the dominant $\alpha$-OMP, accumulating approximately
40\% of the total BMA weight, with the Skyrme-Hartree-Fock-Bogoliubov
level-density model (ld~$=4$) and msl~$=5$ receiving the highest combined
weight. Independently, DEM-3 also exhibits the flattest energy-dependent
shape residuals of all models tested, confirming that its high BMA weight
reflects both a good magnitude fit and the correct energy dependence of
the cross section. 
The energy-dependent factor uncertainty $\mathrm{f.u.}(E) = \exp[\sigma_{\ln,\mathrm{BMA}}(E)]$~\cite{Longland2010},
where $\sigma_{\ln,\mathrm{BMA}}(E)$ is the BMA-weighted standard deviation of $\ln\sigma$
across the model ensemble, was computed in log-space.
The total factor uncertainty on the stellar rate combines the BMA log-space
dispersion $\sigma_{\ln,\mathrm{BMA}}(E)$ and the experimental cross-section
uncertainty $\sigma_{\ln,\mathrm{exp}}(E)$ in quadrature,
\begin{equation}
\sigma_{\ln,\mathrm{tot}} = \sqrt{\sigma_{\ln,\mathrm{BMA}}^2 + \sigma_{\ln,\mathrm{exp}}^2},
\end{equation}
and the corresponding factor uncertainty $\exp[\sigma_{\ln,\mathrm{tot}}(E)]$
defines the uncertainty band.

\subsection{Optimization of the Alpha Optical Model Potential}
\label{sec:geom_opt}

\begin{figure*}[!t]
    \centering
    \includegraphics[width=\textwidth]{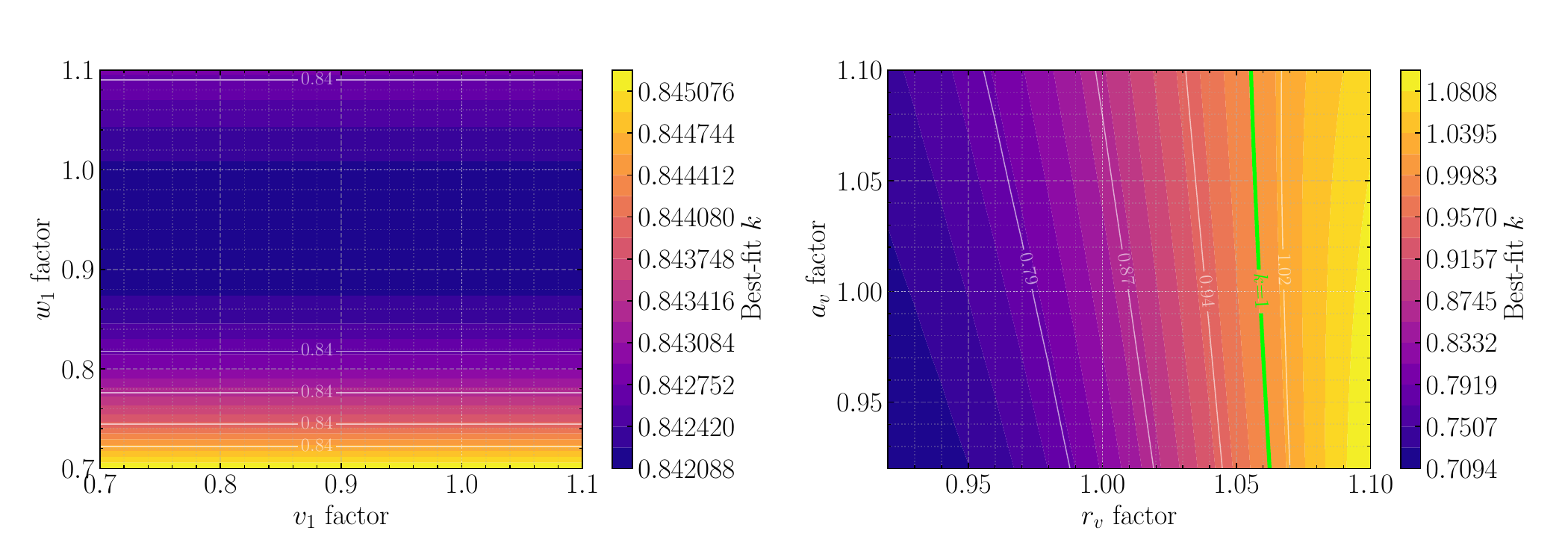}
    \caption{Best-fit scaling factor $k$ as a function of the alpha optical model 
    potential geometry parameters. Left panel: depth grid ($v_1$, $w_1$). 
    Right panel: geometry grid ($r_v$, $a_v$). The lime contour marks $k = 1$.}
    \label{fig:k_landscape}
\end{figure*}

With DEM-3 identified as the preferred potential from the BMA analysis, its geometry parameters were optimized using a two-dimensional grid of the imaginary volume potential via the radius $r_v$ and diffuseness $a_v$ values, keeping ld = 4 and msl = 5. At each grid point, the best-fit normalization factor $k$ was determined analytically from a weighted least-squares minimization. The results are shown in Fig.~\ref{fig:k_landscape} (right panel): $k$ varies smoothly across the grid, and the contour $k = 1$ is satisfied at an adjustment of $r_v$ by $+6$\% and of $a_v$ by $-6$\%. A separate scan over the real and imaginary potential depths $v_1$ and $w_1$ (left panel) shows that $k$ is nearly independent of these parameters.
These parameters for $r_v$ and $a_v$ in combination with the above ld = 4 and msl = 5 were adopted for all subsequent calculations and rate derivations, as it reproduces the measured cross sections without any global renormalization. For completeness we note that the DEM-3 potential includes a dispersive coupling between the real and imaginary part of the potential. Thus, the modification of the parameters $r_v$ and $a_v$ of the imaginary volume potential also affects the real potential, and finally the changes in the calculated cross sections result from both, the direct variations of the imaginary potential via $r_v$ and $a_v$ and the indirect variations of the real potential via the dispersive coupling.

\subsection{Energy Dependence of the Cross Section}

\begin{figure}
    \centering
    \includegraphics[width=0.9\linewidth]{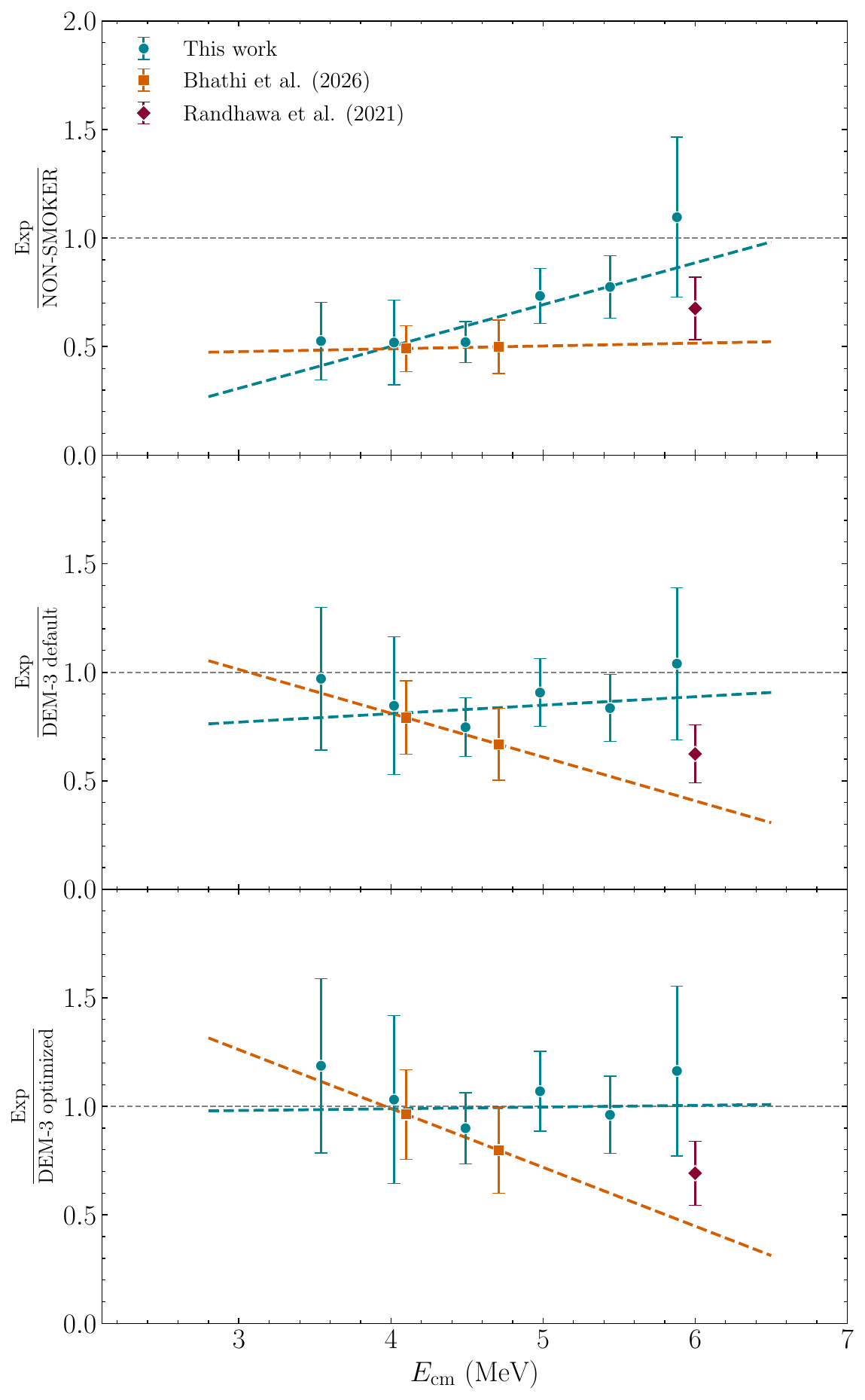}
    \caption{ Ratio of measured $^{59}$Cu$(p,\alpha)^{56}$Ni cross sections to statistical-model predictions as a function of $E_\mathrm{cm}$, for the present work (teal circles), Bhathi {\it{et al.}}\ \cite{Bhathi25} (orange squares), and Randhawa {\it{et al.}}\ \cite{Jaspreet59Cu} (maroon diamond). From top to bottom: NON-SMOKER, DEM-3 default geometry, and DEM-3 with optimized geometry. Dashed lines show linear fits to guide the eye. The flat ratio near unity in the bottom panel confirms that the optimized DEM-3 geometry correctly reproduces the energy dependence of the data.}
    \label{fig:energydependence}
\end{figure}

The ratio of the measured cross sections from Randhawa {\it{et al.}}\ \cite{Jaspreet59Cu}, Bhathi {\it{et al.}}\ \cite{Bhathi25}, and the present work to the statistical-model predictions is shown in Fig.~\ref{fig:energydependence}. A flat ratio at unity indicates that the model correctly reproduces the energy dependence of the data, while a tilted or structured ratio reveals a shape mismatch that no global normalization can remove. The NON-SMOKER prediction systematically overestimates the cross section at low energies and underestimates it at high energies, producing a pronounced tilt. The default DEM-3 geometry improves the situation, but retains a small residual slope. The geometry-optimized calculation yields the flattest ratio, consistent with unity across the full measured range within experimental uncertainties. Reproducing the correct energy dependence is critical for the extrapolation to lower energies: a model with the wrong slope accumulates a systematic error that grows with the distance of the extrapolation and directly biases the derived reaction rate in the astrophysically relevant Gamow window.

\subsection{S-factor behavior}
\label{sec:sfact}
\begin{figure}[!ht]
    \centering
    \includegraphics[width=0.9\linewidth]{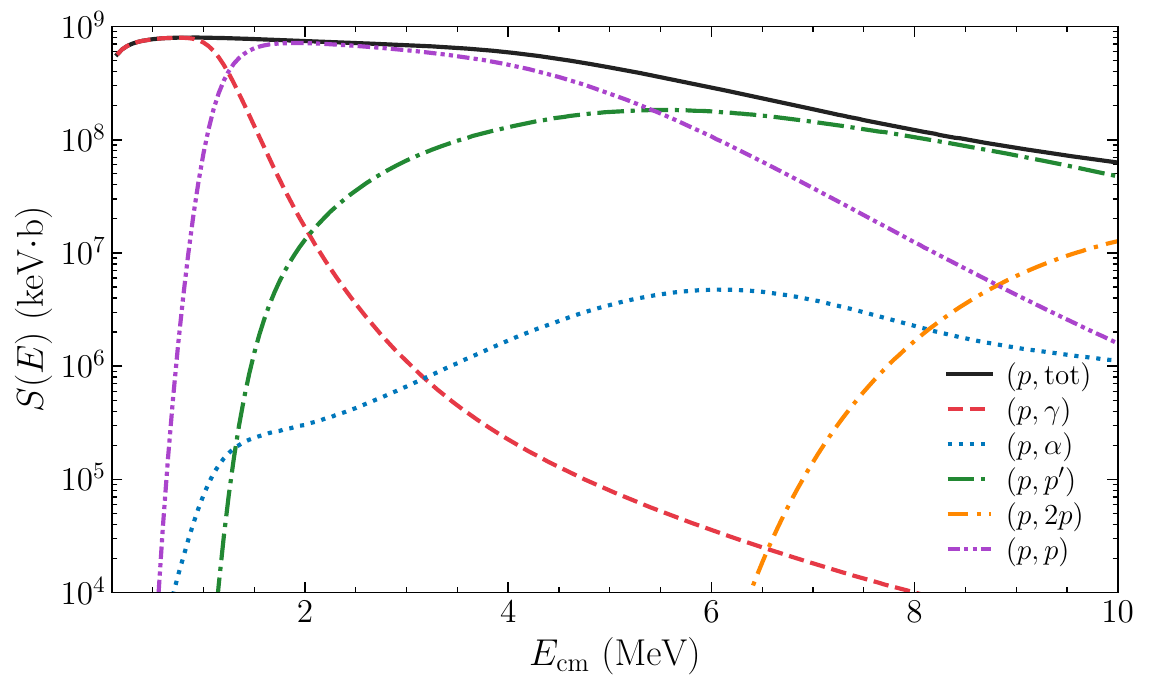}
\caption{Astrophysical S-factors for the $^{59}$Cu$(p,x)$ reactions 
calculated with the DEM-3 potential as a function of $E_{\mathrm{c.m.}}$. 
The total S-factor (solid near-black line), the $(p,\gamma)$ contribution 
(dashed red line), and the $(p,p)$ channel (pink dash-dot-dot line) 
decrease with increasing energy. The $(p,\alpha)$ S-factor (dotted blue 
line) increases with energy and is strongly suppressed at low energies. 
The $(p,p')$ (green dash-dot line) and $(p,2p)$ (orange long-dashed line) channels open at higher energies.}

    \label{fig:Sfactor}
\end{figure}

The astrophysical S-factor, $S(E)$, is defined to remove the dominant Coulomb-barrier effects from the cross section and thus primarily contains the nuclear effects.
Fig.~\ref{fig:Sfactor} shows the S-factors of the different exit channels of the $^{59}$Cu+$p$ system. This presentation allows further interpretation of the calculations discussed in the Sec.~\ref{sec:TALYS}. At the energies under experimental study, the total cross section, $\sigma_{\text{tot}}$, is essentially composed of the compound-elastic $(p,p)$ and compound-inelastic $(p,p')$ contributions, whereas the $(p,\alpha)$ contribution remains below 1\% of $\sigma_{\text{tot}}$. At low energies, below about 3--4 MeV, where the compound-elastic cross section dominates, Eq.~(\ref{eq:trans}) further reduces to $\sigma_{(p,\alpha)} \propto T_{\alpha 0}$ because $T_p \approx T_{p0}$ in this energy region.

Interestingly, Fig.~\ref{fig:Sfactor} shows that the energy dependence of the S-factor varies significantly for the different $(p,x)$ channels. In the definition of the astrophysical S-factor, it is assumed that the cross section is governed by the Coulomb repulsion between point-like charges. Thus, for charged-particle reactions of heavy target nuclei with finite size, the S-factor slightly overcompensates the energy dependence of the cross section, leading to a slightly negative slope of the S-factor with energy. This typical behavior is found for the total cross section \stot\ of $^{59}$Cu+$p$, but the $^{59}\mathrm{Cu}(p,\alpha){}^{56}\mathrm{Ni}$ reaction shows an unusual increasing S-factor with positive slope. This finding can be explained by the fact that the cross section of the $^{59}\mathrm{Cu}(p,\alpha){}^{56}\mathrm{Ni}$ reaction is not only governed by the Coulomb barrier in the entrance channel; in addition, there is further suppression of the cross section towards low energies from the Coulomb barrier in the exit channel. This has a number of interesting consequences, which will be briefly discussed in the following.

Often, the astrophysically most relevant energy window, the Gamow window, is estimated  using an approximate formula:
\begin{equation}
    E_0 = 0.1220 \times (Z_P^2 Z_T^2 A_{\rm{red}} T_9^2)^{1/3}\, {\rm{MeV}} 
    \label{E0}
\end{equation}

which is based on the assumption of a constant (energy-independent) S-factor. The typically negative slope of the S-factor leads to a shift of the most effective energy $E_0$ to lower energies by about $0.5 - 2$ MeV~\cite{Rauscher_PRC2010_gamow}; i.e., experimental data have to reach even lower energies than estimated from the standard formula to cover the Gamow window at a given temperature. Contrary, in the present case the positive slope of the S-factor leads to a shift of the most relevant energy to higher energies. E.g., at $T_9 = 3$, $E_0 = 2382$ keV from the simple approximation formula in Eq.~(\ref{E0}), corresponding to an excitation energy in the $^{60}$Zn compound nucleus of $E^\ast$ = 7487 keV. The real energy window is located about 500 keV higher. \par Table~\ref{tab:gamow_windows} shows the Gamow peaks calculated numerically using the S factor of the $(p,\alpha)$ reaction shown in Fig.~\ref{fig:Sfactor}, $\tilde{E}_0$ (see Ref.~\cite{Rauscher_PRC2010_gamow}), in comparison with the Gamow peak obtained from the approximation formula, $E_0$ in Eq.~(\ref{E0}), and the shift $\delta = \tilde{E}_0 - E_0$ between the two for different temperatures.

\begin{table}[htb]
\centering
\caption{Gamow peak energies $\tilde{E}_0$ calculated using the S factor, $E_0$ calculated with the approximation formula, and the shift $\delta = \tilde{E}_0 - E_0$ for the $^{59}$Cu($p,\alpha$) reaction.}
\begin{tabular}{cccc}
\toprule
$T_9$  & $\tilde{E}_0$ [MeV]  & $E_0$ [MeV]  & $\delta$ [MeV] \\
\midrule
0.5   & 0.9 & 0.72 & 0.18 \\
1.0   & 1.3 & 1.14 & 0.16 \\
1.5   & 1.6 & 1.49 & 0.11 \\
2.0   & 1.95 & 1.81 & 0.14 \\
2.5   & 2.35 & 2.1 & 0.25 \\
3.0   & 2.85 & 2.38 & 0.47 \\
\hline
\end{tabular}
\label{tab:gamow_windows}
\end{table}
\begin{figure}[htbp]
    \centering
    \includegraphics[width=0.9\linewidth]{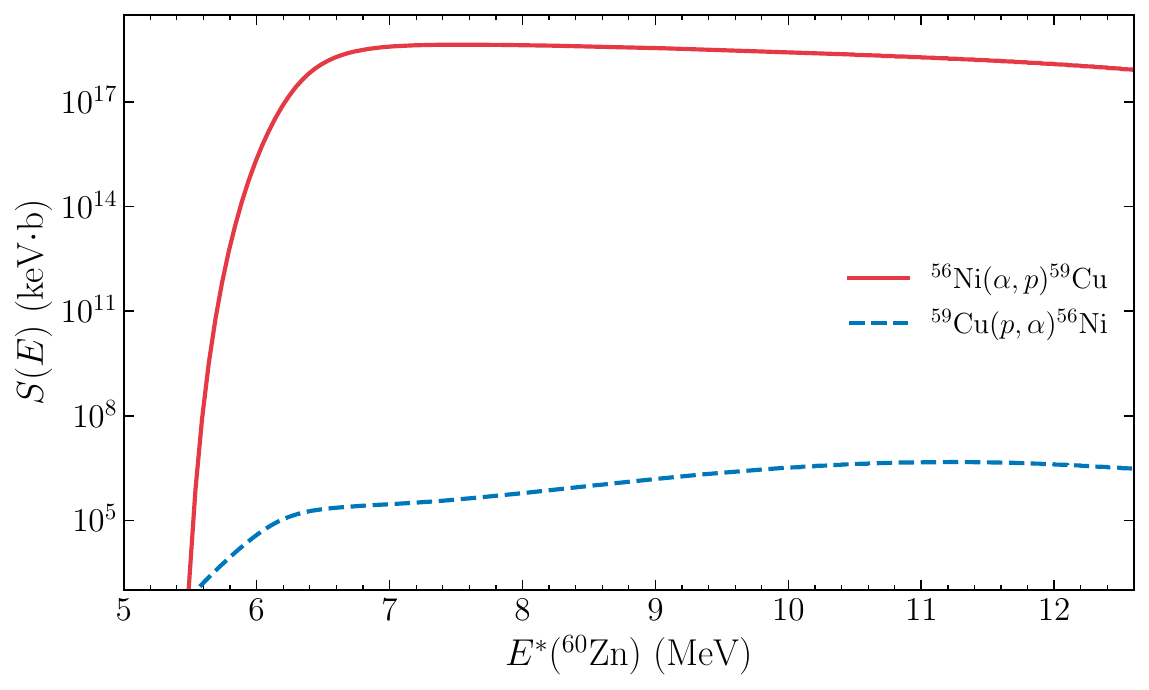}
\caption{Astrophysical S-factors for the $^{56}\mathrm{Ni}(\alpha,p)^{59}\mathrm{Cu}$ and $^{59}\mathrm{Cu}(p,\alpha)^{56}\mathrm{Ni}$ reactions plotted as a function of the excitation energy of the compound nucleus $^{60}\mathrm{Zn}$.}
    \label{fig:ap_pa_comp}
\end{figure}

For completeness, we note that the S-factor of the reverse $^{56}\mathrm{Ni}(\alpha,p)^{59}\mathrm{Cu}$ reaction exhibits the conventional negative slope once the energy is sufficiently above the reaction threshold at $Q = 2.41$~MeV, as illustrated in Fig.~\ref{fig:ap_pa_comp}. Using the Gamow-peak approximation, the most effective energy for the $^{56}$Ni+$\alpha$ reaction at $T_9 = 3$ is $E_0 = 5762$~keV, corresponding to an excitation energy of $E^\ast = 8454$~keV in $^{60}$Zn. While a downward shift of the astrophysical relevant energy window is generally expected for a decreasing S-factor, in the present case this shift amounts to about 500~keV at $T_9 = 3$.\par
The stellar reaction rates of the $^{59}\mathrm{Cu}(p,\alpha){}^{56}\mathrm{Ni}$ forward and $^{56}\mathrm{Ni}(\alpha,p){}^{59}\mathrm{Cu}$ reverse reactions are directly linked by detailed balance. This means that the forward and reverse rates must proceed at the same excitation energy in the $^{60}$Zn compound nucleus. At $T_9 = 3$, the most relevant excitation energy in $^{60}$Zn is around $E^\ast = 8$ MeV, in between the two above estimates from the simple formula of 7487 keV for the $^{59}$Cu + p channel and 8454 keV for the $^{56}$Ni + $\alpha$ channel.

Therefore, using the standard formula and the S factor of the (p,$\alpha$) reaction shown in Fig.~\ref{fig:Sfactor}, we estimate that our experimental data cover a temperature range of $T_9 \sim 2.6$--10.5.

\subsection{Reaction Rate calculation }

The present work gives an experimentally constrained reaction rate of the ground state contribution of $^{59}$Cu to the total stellar rate in the  $T_9\sim$ 2.6-10.5 range. To extend the rate toward lower temperatures, the total cross sections measured here were combined with the extrapolated TALYS cross sections. The recommended stellar reaction rates for $^{59}$Cu(p,$\alpha$)$^{56}$Ni obtained in this work are shown in Table~\ref{table_rr}. 
\begin{table}[!ht]
\centering
\caption{Recommended stellar reaction rates for $^{59}$Cu(p,$\alpha$)$^{56}$Ni. Rate written in [cm$^3$ mol$^{-1}$ s$^{-1}$] units}
\begin{tabular}{cccc}
\toprule
$T_9$  & Low  & Recommended  & High  \\
\midrule
0.05  & 6.828$\times10^{-37}$ & 1.111$\times10^{-36}$ & 1.809$\times10^{-36}$ \\
0.10  & 9.985$\times10^{-28}$ & 1.625$\times10^{-27}$ & 2.645$\times10^{-27}$ \\
0.15  & 8.765$\times10^{-23}$ & 1.426$\times10^{-22}$ & 2.322$\times10^{-22}$ \\
0.20  & 1.484$\times10^{-19}$ & 2.416$\times10^{-19}$ & 3.932$\times10^{-19}$ \\
0.25  & 3.318$\times10^{-17}$ & 5.401$\times10^{-17}$ & 8.790$\times10^{-17}$ \\
0.30  & 2.253$\times10^{-15}$ & 3.666$\times10^{-15}$ & 5.967$\times10^{-15}$ \\
0.40  & 1.241$\times10^{-12}$ & 2.020$\times10^{-12}$ & 3.287$\times10^{-12}$ \\
0.50  & 1.169$\times10^{-10}$ & 1.902$\times10^{-10}$ & 3.096$\times10^{-10}$ \\
0.60  & 3.631$\times10^{-9}$  & 5.910$\times10^{-9}$  & 9.618$\times10^{-9}$  \\
0.70  & 5.383$\times10^{-8}$  & 8.761$\times10^{-8}$  & 1.426$\times10^{-7}$  \\
0.80  & 4.773$\times10^{-7}$  & 7.767$\times10^{-7}$  & 1.264$\times10^{-6}$  \\
0.90  & 2.965$\times10^{-6}$  & 4.760$\times10^{-6}$  & 7.642$\times10^{-6}$  \\
1.00  & 1.404$\times10^{-5}$  & 2.218$\times10^{-5}$  & 3.504$\times10^{-5}$  \\
1.50  & 2.985$\times10^{-3}$  & 4.390$\times10^{-3}$  & 6.456$\times10^{-3}$  \\
2.00  & 8.283$\times10^{-2}$  & 1.147$\times10^{-1}$  & 1.590$\times10^{-1}$  \\
2.50  & 9.191$\times10^{-1}$  & 1.212$\times10^{0}$   & 1.598$\times10^{0}$   \\
3.00  & 6.184$\times10^{0}$   & 7.803$\times10^{0}$   & 9.846$\times10^{0}$   \\
3.50  & 2.990$\times10^{1}$   & 3.637$\times10^{1}$   & 4.425$\times10^{1}$   \\
4.00  & 1.126$\times10^{2}$   & 1.333$\times10^{2}$   & 1.577$\times10^{2}$   \\
5.00  & 9.037$\times10^{2}$   & 1.041$\times10^{3}$   & 1.200$\times10^{3}$   \\
6.00  & 4.390$\times10^{3}$   & 4.815$\times10^{3}$   & 5.282$\times10^{3}$   \\
7.00  & 1.484$\times10^{4}$   & 1.550$\times10^{4}$   & 1.620$\times10^{4}$   \\
8.00  & 3.802$\times10^{4}$   & 3.868$\times10^{4}$   & 3.934$\times10^{4}$   \\
9.00  & 7.832$\times10^{4}$   & 8.032$\times10^{4}$   & 8.238$\times10^{4}$   \\
10.00 & 1.386$\times10^{5}$   & 1.457$\times10^{5}$   & 1.532$\times10^{5}$   \\
\bottomrule
\end{tabular}
\label{table_rr}
\end{table}

Following Rauscher~\cite{Rauscher_2011_gs_constrain}, the true ground--state contribution to the stellar reaction rate is given by
\begin{equation}
X(T) 
= \frac{w_0(T)\,R_0(T)}{R^{*}(T)}
\label{eq:X0_def}
\end{equation}
where $R_0$ is the reaction rate on the ground state of the target nucleus and $R^{*}$ is the full stellar reaction rate including transitions from thermally populated excited states. The factor
\begin{equation}
w_0(T) = \frac{(2J_0+1)}{G(T)}
= \frac{1}{G_0(T)}
\end{equation}
describes the thermal population of the ground state, with $J_0$ the ground--state spin and $G(T)$ the nuclear partition function normalized by $G_0(T) = G(T)/(2J_0+1)$.

Using the stellar enhancement factor,
\begin{equation}
f_{\mathrm{SEF}}(T) = \frac{R^{*}(T)}{R_0(T)}
\label{eq:fsef_def}
\end{equation}

\begin{figure}[!htb]
    \centering
    \includegraphics[width=0.9\linewidth]{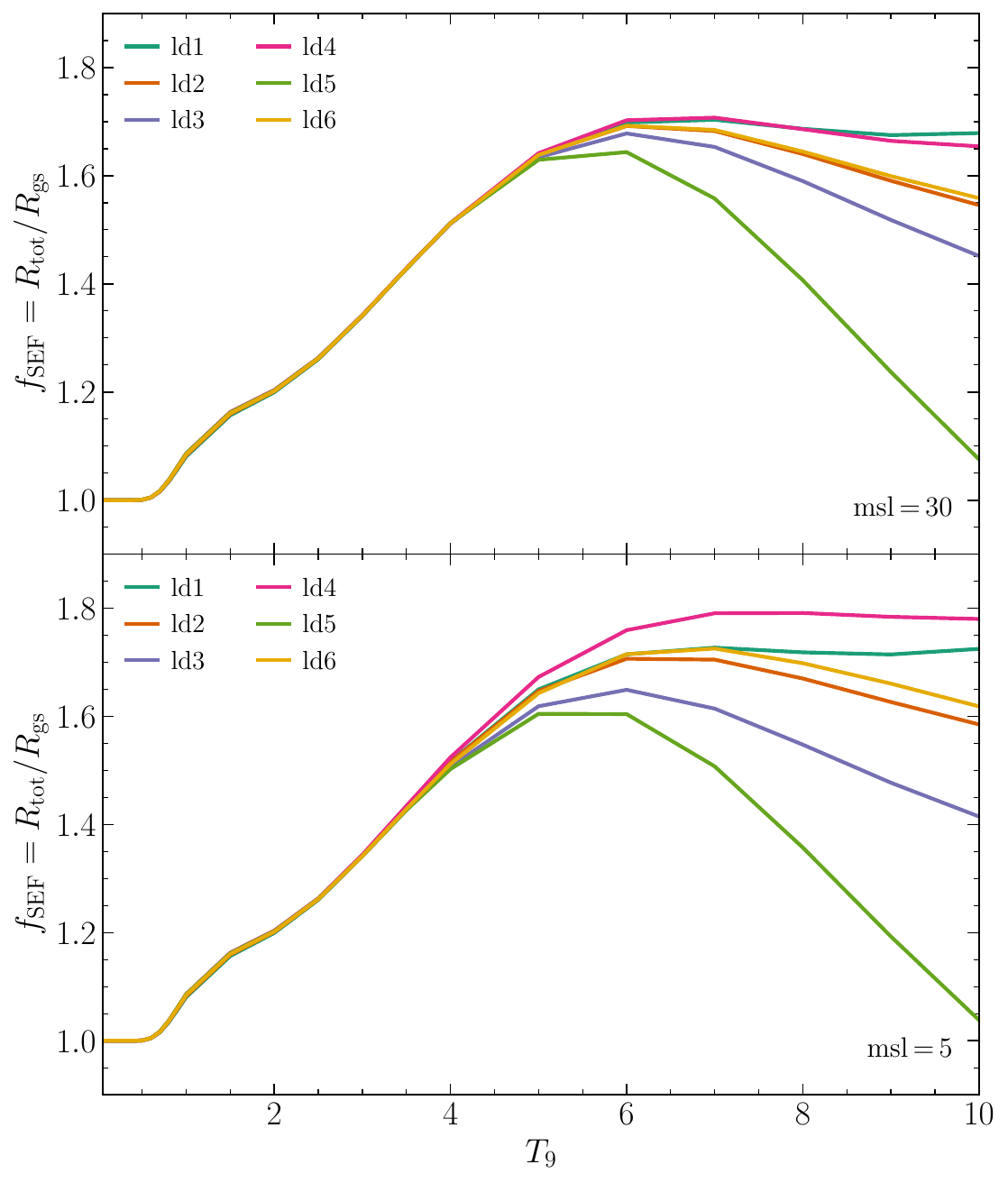}
\caption{
Stellar enhancement factor $f_{\mathrm{SEF}}(T)$ from TALYS, illustrating the spread from varying ldmodel and maxlevelstar.}
\label{fig:sef_models}
\end{figure}

 Eq.~(\ref{eq:X0_def}) becomes
\begin{equation}
X(T)
= \frac{1}{f_{\mathrm{SEF}}(T) G_0(T)}
\label{eq:X0_final}
\end{equation}
This quantity determines the maximal fraction of the stellar reaction rate that can be constrained by a laboratory measurement performed on target nuclei in their ground state.
 To obtain $X(T)$, first, $f_{\mathrm{SEF}}(T)$ was extracted by calculating the ratio of the stellar rate to the rate derived in this work using the TALYS parameters that best reproduce the measured cross section: ld = 4 (Skyrme-Hartree-Fock-Bogolyubov), msl = 5, and the optimized dispersive $\alpha$-OMP DEM-3. Using the TALYS calculations, where nuclear partition functions are derived from nuclear level-density estimates within the combinatorial model of Goriely \textit{et al.}~\cite{Goriely_G}, we extracted the temperature-dependent partition function $G(T)$ of $^{59}$Cu and subsequently determined the ground-state contribution $X(T)$. 
 To evaluate the uncertainties on the calculated $f_{\mathrm{SEF}}(T)$, the level-density models (ld = 1--6) and the number of included discrete states in $^{59}$Cu (msl = 5 and 30) were varied, using the DEM-3 potential. The resulting spread in $f_{\mathrm{SEF}}(T)$ is illustrated in Fig.~\ref{fig:sef_models}. The results demonstrate that the first few excitated states in $^{59}$Cu dominate $f_{\mathrm{SEF}}(T)$. In Fig.~\ref{fig:sef_models}  it can be seen that all level-density models yield nearly identical values independently of the amount of discrete states implemented up to $T_{9}\approx4$. At higher temperatures the spread between different ld and msl combinations remains moderate, with ld = 5 corresponding to the Skyrme–Hartree–Fock–Bogolyubov  combinatorial level-density from numerical tables model, yielding the lowest values of $f_{\mathrm{SEF}}(T)$. The spread in $f_{\mathrm{SEF}}(T)$ resulting from the variations in ld and msl was quantified by the uncertainty factor defined as:

\begin{equation}
\hspace*{3em}
f.u._{\mathrm{SEF}}(T) = \sqrt{\frac{f_{\mathrm{SEF}}^{\max}(T)}{f_{\mathrm{SEF}}^{\min}(T)}} \
\label{eq:fusef}
\end{equation}

\begin{figure}[!htb]
    \centering
    \includegraphics[width=0.9\linewidth]{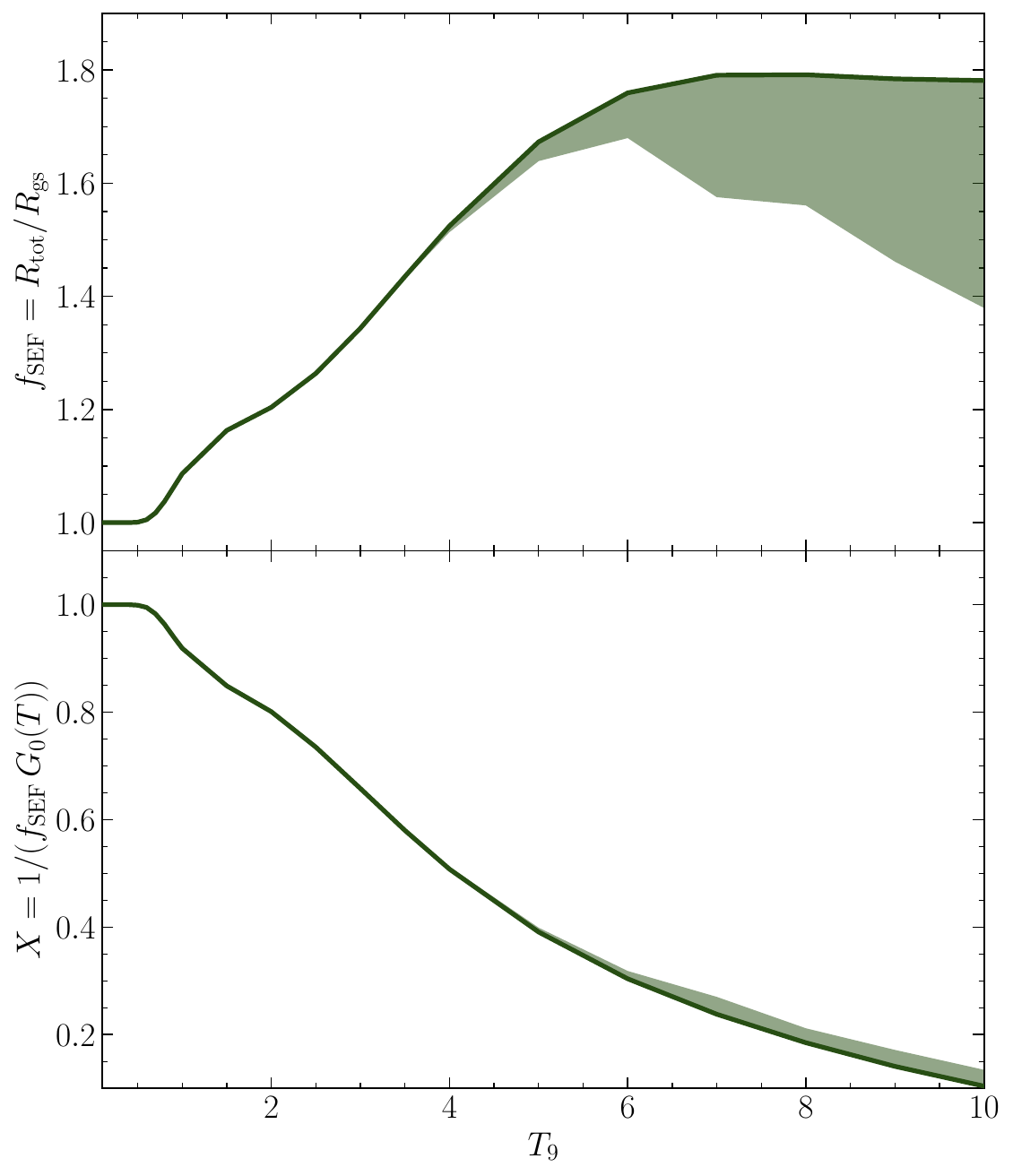}
    \caption{Stellar enhancement factor (top) and ground-state contribution (bottom) for the
$^{59}$Cu(p,$\alpha$)$^{56}$Ni reaction as a function of temperature in units of $T_{9}$. }
\label{fig:gs_fraction}
\end{figure}

This spread is shown as the shaded band in the upper panel of  Fig.~\ref{fig:gs_fraction}. 
 
The corresponding uncertainty in the ground-state contribution, $X(T)$, obtained by propagating this factor, is displayed in the lower panel.

In Fig.~\ref{fig:gs_fraction}, the ground-state contribution $X(T)$ decreases smoothly with increasing temperature, as expected from the growing population of thermally excited states. The uncertainty in $X(T)$, propagated from $f.u._{\mathrm{SEF}}(T)$ for the selected ld, remains small over the full temperature range, indicating that the SEF-related uncertainty has no significant impact on the ground-state contribution; the nuclear partition function $G_0(T)$ was kept fixed in this propagation.

 As shown by Rauscher et al~\cite{Rauscher_2011_gs_constrain}, the ground-state contribution  sets the maximum possible uncertainty reduction for a measured laboratory cross section.  In the temperature range covered by our measurement, $2.6 \leq T_{9} \leq 10.5$, we obtain $X \simeq 0.75$ at $T_{9} \approx 2.6$ decreasing to $X \simeq 0.10$ at $T_{9} \gtrsim 10$. This implies that our experiment directly constrains approximately $10$--$75\%$ of the stellar reaction rate in this temperature range, while the remaining  originates from reactions on thermally populated excited states and therefore remains model dependent. \par

\begin{figure}[!ht]
    \centering
    \includegraphics[width=0.9\linewidth]{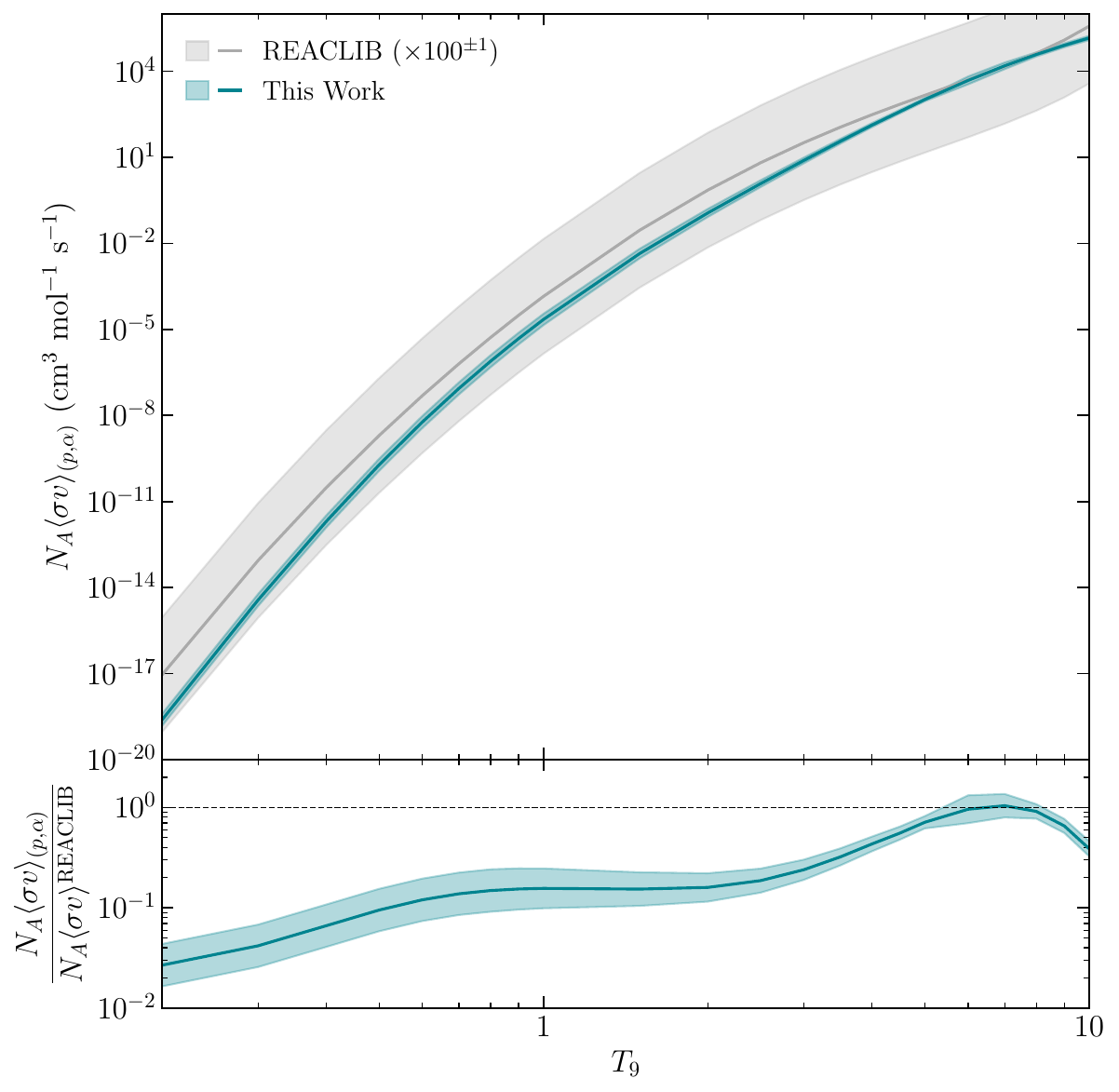}
    \caption{(Top) Recommended $^{59}\mathrm{Cu}(p,\alpha)^{56}\mathrm{Ni}$ stellar reaction rate from the present work (teal line) with propagated uncertainties (shaded band) and the REACLIB  rate from NON\textendash SMOKER calculations~\cite{ths8} (black line). (Bottom) Logarithm of the ratio of our recommended stellar rate to the REACLIB rate.}
    \label{fig:rate}
\end{figure}

The reaction rate obtained in this work is compared to the widely used REACLIB rate~\cite{ths8} in Fig.~\ref{fig:rate} over the temperature range $0.1 \leq T_9 \leq 10$. As seen in the figure, our rate is consistently lower than REACLIB, even when uncertainties are considered. The only exception occurs in the temperature range $T_9 \approx 4.55$--9, where our rate exceeds REACLIB, mainly due to a pronounced dip in the REACLIB parametrization.

In REACLIB, thermonuclear reaction rates are expressed as analytic fits using a seven-parameter formalism that describes their temperature dependence, following Ref.~\cite{ths8}:
\begin{equation}
\begin{split}
N_A \langle \sigma v \rangle_{\mathrm{REACLIB}} = 
\exp\Bigl(
a_0
+ \frac{a_1}{T_9}
+ \frac{a_2}{T_9^{1/3}}
+ a_3 T_9^{1/3} \\
\hspace{1.25cm}
+\, a_4 T_9
+ a_5 T_9^{5/3}
+ a_6 \ln T_9
\Bigr)
\end{split}
\label{eq:reaclib_param}
\end{equation}

The REACLIB coefficients adopted in this work for the $^{59}$Cu$(p,\alpha)^{56}$Ni and $^{59}$Cu$(p,\gamma)^{60}$Zn reactions are listed in Table~\ref{tab:reaclib_coeffs}.

\begin{table}[!h]
\centering
\caption{REACLIB parameters for the 
$^{59}\mathrm{Cu}(p,\alpha)^{56}\mathrm{Ni}$ and $^{59}\mathrm{Cu}(p,\gamma)^{60}\mathrm{Zn}$ reactions extracted from Ref.~\cite{ths8}.}
\begin{tabular}{c r r}
\toprule
Parametrization & REACLIB &  REACLIB \\
 Coefficients&  $^{59}\mathrm{Cu}(p,\alpha)^{56}\mathrm{Ni}$  & $^{59}\mathrm{Cu}(p,\gamma)^{60}\mathrm{Zn}$\\
\midrule
$a_0$ &$ 2.076880 \times 10^{1}$ & $ 3.722600 \times 10^{1}$ \\
$a_1$ &$ 0 $ & $ 0 $ \\
$a_2$ &$ -3.998080 \times 10^{1}$ & $ -3.998080 \times 10^{1}$ \\
$a_3$ &$ 1.367730 \times 10^{1}$ & $ 1.173110 \times 10^{0}$ \\
$a_4$ &$ -3.764290$ & $ -2.904860 \times 10^{0}$ \\
$a_5$ &$ 4.380960 \times 10^{-1}$ & $ 3.396440 \times 10^{-1}$ \\
$a_6$  &$ -6.666670 \times 10^{-1}$ & $ -6.666670 \times 10^{-1}$ \\
\bottomrule
\end{tabular}
\label{tab:reaclib_coeffs}
\end{table}

\section{Comparison of Reaction Rates}\label{sec:comparison}

 To evaluate the impact of our newly calculated stellar reaction rate, we compare it with previously derived rates as well as with rates from commonly used reaction-rate libraries, such as REACLIB and STARLIB. Since the impact of this reaction in XRBs and the $\nu p$-process in CCSNe strongly depends on the ratio of the $(p,\alpha)$ to $(p,\gamma)$ rates, we also review previous work on the $(p,\gamma)$ reaction and compare the two rates. Given that the relevant temperatures for XRBs and the $\nu p$-process are below $T_9 \lesssim 3$, and that previous experimentally determined reaction rates were derived only at low temperatures, we focus our discussion here on $T_9 = 0.2$–3.

\subsection{Comparison of available \texorpdfstring{$^{59}$Cu(p,$\alpha$)$^{56}$Ni}{59Cu(p,a)56Ni} reaction rates}

In addition to the reaction rate obtained from the direct measurement of Bhathi \textit{et al.}~\cite{Bhathi25} at higher energies, indirect constraints based on available experimental information on the nuclear properties of the compound nucleus $^{60}$Zn, combined with shell-model calculations, have been provided by Refs.~\cite{Kim_2022,Lotay}. Since the reaction rate of Ref.~\cite{Lotay} is not yet available, it will not be shown here. The indirect constraint obtained by Kim \textit{et al.}~\cite{Kim_2022} used properties of known resonances of $^{60}$Zn up to 2.8 MeV, and performed shell-model calculations to estimate the $\gamma$-ray partial widths. The $\alpha$ widths were randomly sampled using the Porter–Thomas distribution, and a Monte Carlo technique was then employed to extract the thermonuclear reaction rate over the temperature range $T_9 = 0.1$–2.

\begin{figure}[!htb]
\centering
\includegraphics[width=0.9\linewidth]{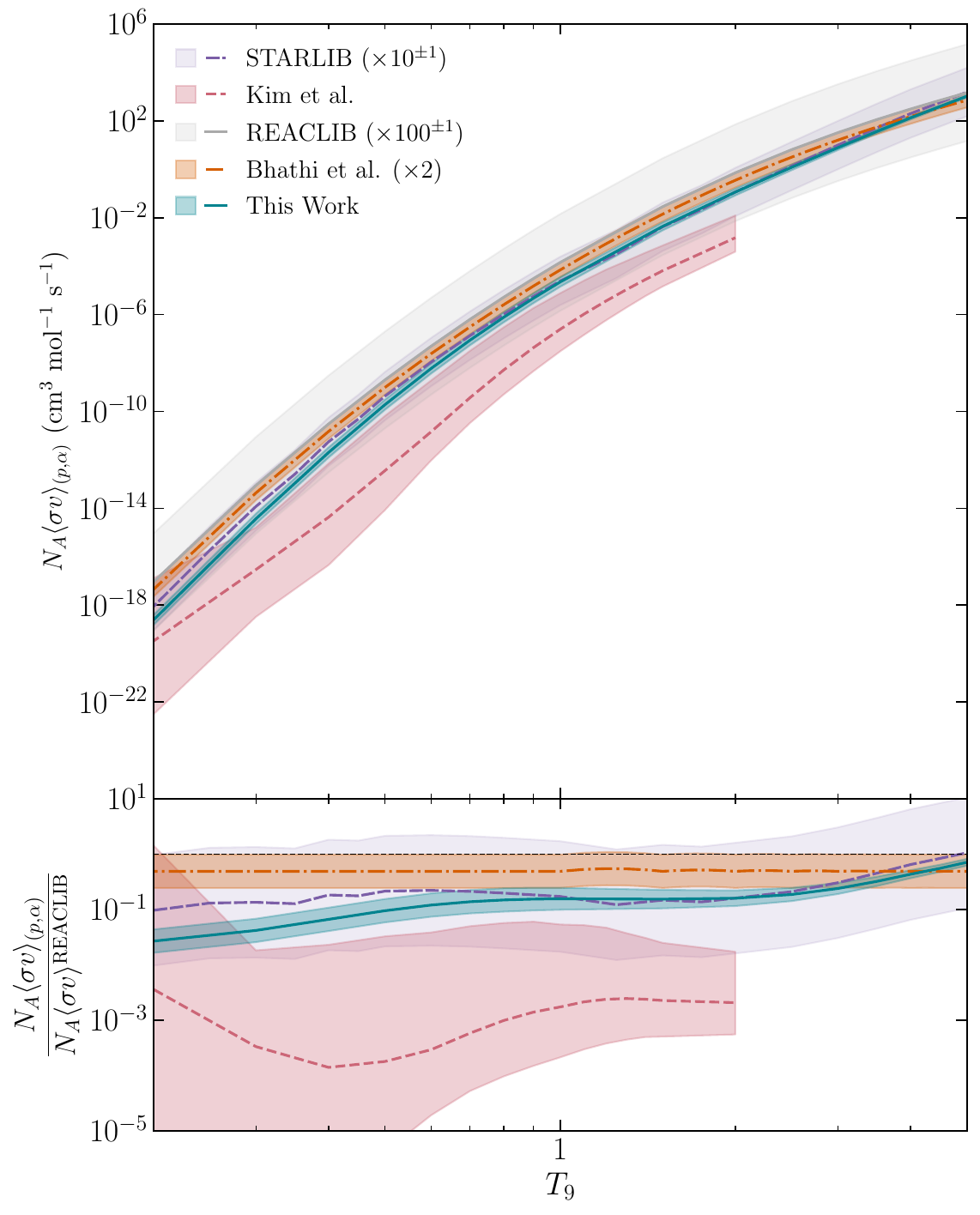}
\caption{Comparison of stellar reaction rates for $^{59}\mathrm{Cu}(p,\alpha){}^{56}\mathrm{Ni}$.
The top panel shows the median rate from this work  with its uncertainty band, together with literature evaluations: Bhathi \textit{et al.}~\cite{Bhathi25}, Kim \textit{et al.}~\cite{Kim_2022}, the REACLIB evaluation, and the STARLIB rate~\cite{Sallaska_2013}. The bottom panel shows the corresponding ratios of each rate to the REACLIB evaluation; the horizontal dashed line indicates unity.}
\label{fig:pa_rate_ratio}
\end{figure}

Fig.\ref{fig:pa_rate_ratio} shows the reaction rate $N_A\langle\sigma v\rangle$ for the $^{59}$Cu$(p,\alpha)$ reaction up to $T_9 = 5$, comparing the present result with previous measurements by Bhathi \textit{et al.}\cite{Bhathi25}, Kim \textit{et al.}\cite{Kim_2022} and commonly used rate evaluations such as REACLIB and STARLIB\cite{Sallaska_2013}. The upper panel displays the absolute rates, while the lower panel shows the ratios relative to REACLIB. The present rate lies systematically below REACLIB over the full temperature range. It is about 40 times lower at $T_9 = 0.2$ and about a factor of 4 lower at $T_9 = 3$. Compared to STARLIB (which is based on the original DEM-3 $\alpha$-OMP), our rate is nearly identical for temperatures $T_9 \gtrsim 1$, while it is lower by a factor of $\sim$4 at $T_9 = 0.2$. Compared to the experimentally constrained reaction rates, our rate falls between them over the majority of the relevant temperature range. Since the recent constraint from Bhathi \textit{et al.}\cite{Bhathi25} is based on the REACLIB rate scaled by a factor of 0.49, their rate is larger than ours by a factor of $\sim$18 at $T_9 = 0.2$ and about a factor of 2 larger at $T_9 = 3$. As observed in Fig.\ref{fig:xsec_comparison}, NON-SMOKER, which is used for the REACLIB rates scaled by a factor of 0.49, does not reproduce the energy dependence of our data and leads to an overestimation of the reaction rate at low temperatures. Our rate is close to the REACLIB prediction scaled by 0.49 at temperatures around $T_9 \gtrsim 3.5$ (see Fig.\ref{fig:rate}), and therefore consistent with the rate of Ref.\cite{Bhathi25} in the region where the experimental data agree. The rate of Kim \textit{et al.} exhibits a significantly different temperature dependence at low $T_9$, with large discrepancies for $T_9 \gtrsim 0.3$, exceeding three orders of magnitude, primarily due to the use of Porter–Thomas--sampled $\alpha$ widths in the absence of experimental constraints.

Overall, the present work provides a significantly improved and experimentally constrained $^{59}$Cu$(p,\alpha)^{56}$Ni reaction rate over the temperature range relevant to explosive nucleosynthesis.

\subsection{Comparison of \texorpdfstring{$^{59}$Cu$(p,\alpha){}^{56}$Ni}{59Cu(p,a)56Ni} and \texorpdfstring{$^{59}$Cu$(p,\gamma){}^{60}$Zn}{59Cu(p,g)60Zn} Reaction Rates}

Given that the impact of the $(p,\alpha)$ reaction in XRBs and the $\nu p$-process depends sensitively on the competing $(p,\gamma)$ reaction rate, we briefly review the current status of this reaction in this section. As mentioned above, if the $(p,\alpha)$ reaction rate exceeds the $(p,\gamma)$ reaction rate, a strong NiCu  cycle is established, causing the reaction flow to predominantly cycle back to lower masses. In contrast, a smaller $(p,\alpha)$ reaction rate allows nucleosynthesis to proceed toward heavier nuclei.

\begin{figure}[htbp]
\centering
\includegraphics[width=0.9\columnwidth]{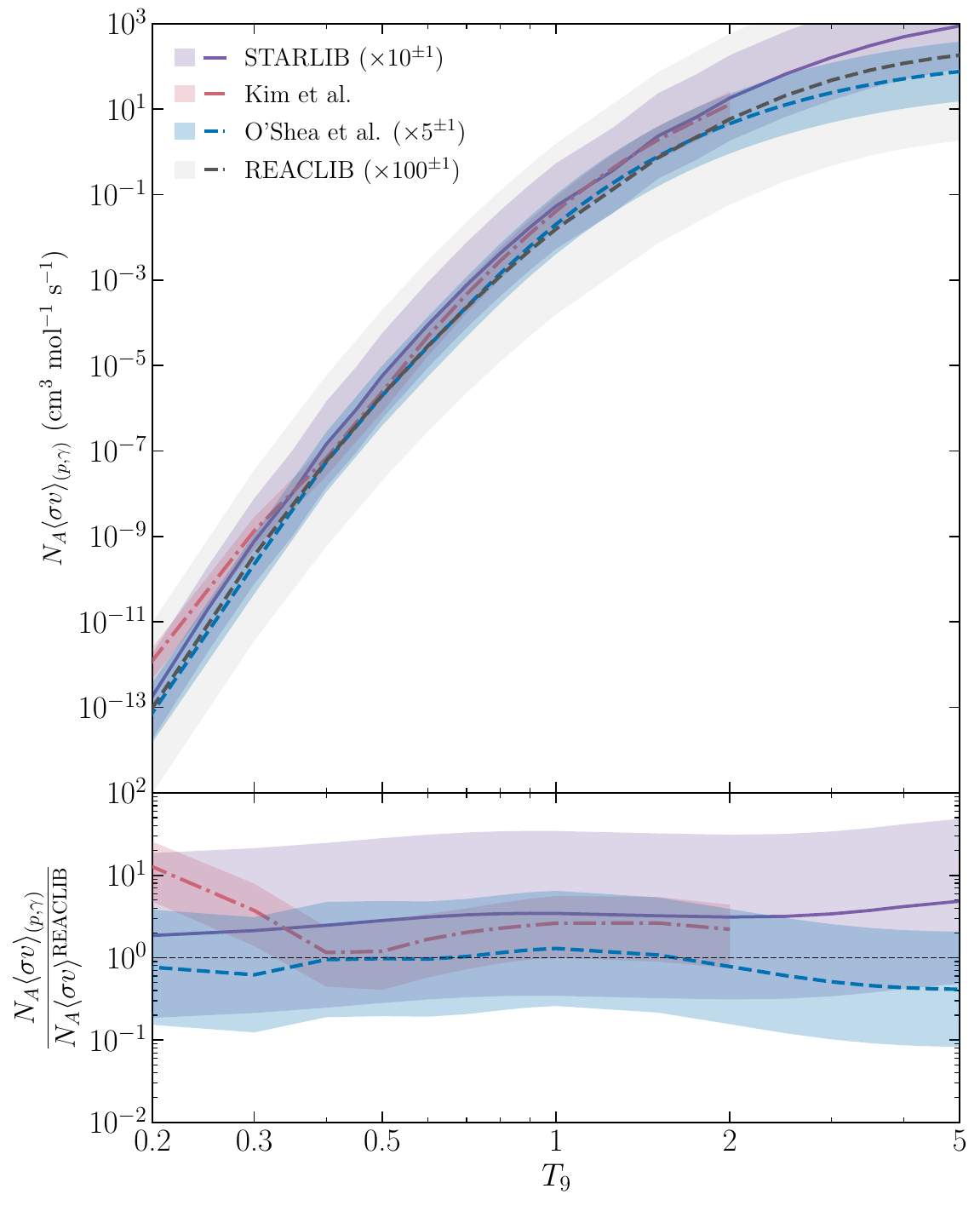}
\caption{Comparison of stellar reaction rates for $^{59}\mathrm{Cu}(p,\gamma){}^{60}\mathrm{Zn}$.
The top panel shows the REACLIB evaluation, the rate from Kim \textit{et al.}~\cite{Kim_2022}, the rate from O’Shea \textit{et al.}~\cite{Lotay}, and the STARLIB rate~\cite{Sallaska_2013}. The bottom panel shows ratios of the literature and library rates to the REACLIB evaluation; the horizontal dashed line indicates unity.}
\label{fig:pg_rate_ratio}
\end{figure}

Since the $(p,\gamma)$ reaction cross section is expected to be small, this reaction has not been measured directly. Similar to what was done for the $(p,\alpha)$ rate, an indirect constraint on the $(p,\gamma)$ reaction was obtained by Kim {\it{et al.}}~\cite{Kim_2022} using available experimental information on the nuclear properties of the compound nucleus $^{60}$Zn, combined with shell-model calculations to estimate the relevant partial widths. A Monte Carlo approach was then used to extract the thermonuclear reaction rate over the temperature range $T_9 = 0.1$–2. They found that the $(p,\gamma)$ rate is higher than the REACLIB rate over the entire temperature range. While the rate exceeds the REACLIB prediction by more than an order of magnitude at low temperatures, the difference decreases to less than a factor of 3 in the temperature range $T_9 = 0.3$–2. More recently, an experimental constraint on this reaction was obtained by O’Shea \textit{et al.}~\cite{Lotay}, who used the $^{59}$Cu$(d,n){}^{60}$Zn reaction to study proton-unbound resonant states in $^{60}$Zn. They identified 15 proton-unbound levels and employed shell-model calculations to estimate the $\gamma$ partial widths. The resulting reaction rate was found to be in good agreement with the REACLIB rate. The uncertainties in the $(p,\gamma)$ reaction rate obtained by Ref.~\cite{Lotay} were found to be $\sim$30\% due to the resonance energies, and a factor of $\sim$2–5 due to the uncertainties in individual resonance strengths; we have used a factor of 5 uncertainty for their measurement. Fig.~\ref{fig:pg_rate_ratio} compares the reaction rates obtained using experimental constraints from Refs.~\cite{Kim_2022,Lotay} with those from the REACLIB and STARLIB reaction-rate libraries. In comparison to REACLIB, the STARLIB rate is approximately a factor of two larger.

Overall, despite remaining discrepancies in the $(p,\gamma)$ reaction rate, the various rate evaluations are consistent to within a factor of 2 for temperatures $T_9 \gtrsim 0.3$. However, now that tighter constraints have been placed on the $(p,\alpha)$ rate, efforts should focus on reducing the large uncertainties in the $(p,\gamma)$ reaction to better constrain the NiCu cycle over the temperature range relevant to XRBs and the $\nu p$-process.

 \begin{figure}[htbp]
\centering
\includegraphics[width=0.9\linewidth]{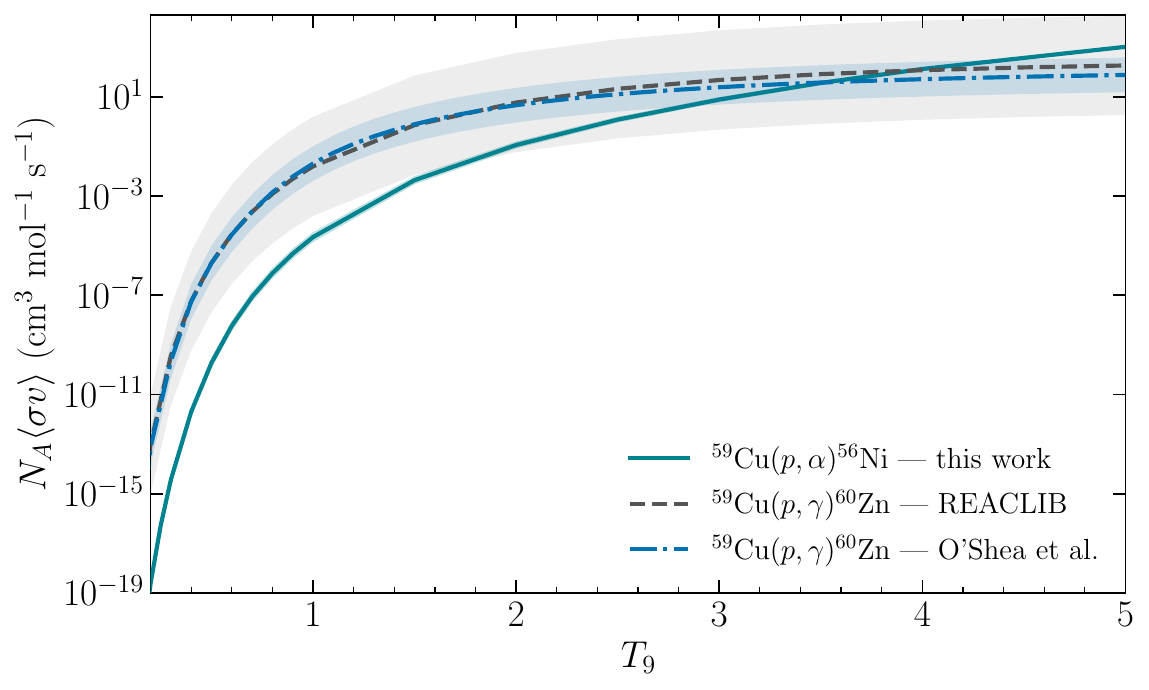}
\caption{Recommended stellar reaction rate for the 
$^{59}\mathrm{Cu}(p,\alpha){}^{56}\mathrm{Ni}$ reaction from this work, 
compared to the $^{59}\mathrm{Cu}(p,\gamma){}^{60}\mathrm{Zn}$ rates from REACLIB and from O'Shea \textit{et al.}\ \cite{Lotay}.}
\label{fig:rate_ratio}
\end{figure}

Fig.~\ref{fig:rate_ratio} shows a comparison, up to 5~GK, of the $(p,\alpha)$ reaction rate derived in this work with the $(p,\gamma)$ rates from REACLIB and O’Shea \textit{et al.}~\cite{Lotay}. As seen in the figure, the $(p,\alpha)$ rate remains consistently below the $(p,\gamma)$ rates across the entire temperature range. This indicates that the strength of the NiCu  cycle is small. Further details on the strength of the NiCu  cycle and its astrophysical impact on XRBs and the $\nu p$-process will be discussed in Ref.~\cite{PRL}.

\section{Conclusions}\label{sec:conclusions}

We have presented a new, experimentally constrained determination of the stellar $^{59}$Cu$(p,\alpha)^{56}$Ni reaction rate based on a direct measurement of the excitation function using the MUSIC active-target detector at FRIB. The present measurement provides angle- and energy-integrated cross sections over the center-of-mass energy range $E_{\mathrm{cm}} = 2.43$--5.88~MeV, extending to lower energies than previously measured.

Statistical-model calculations using TALYS were performed to extrapolate the cross sections to lower energies and to extract the stellar reaction rate at temperatures relevant for XRBs and the $\nu p$-process in CCSNe. To robustly constrain this extrapolation, we performed a systematic optimization of the DEM-3 $\alpha$-optical model potential geometry to best reproduce the measured cross sections. Furthermore, to rigorously quantify the model-selection uncertainty, we conducted a Bayesian model averaging (BMA) analysis over 96 TALYS parameter combinations, incorporating systematic variations in the $\alpha$-OMP, nuclear level densities, and the treatment of discrete levels. Through this BMA framework, the resulting recommended stellar rate carries a fully propagated, temperature-dependent uncertainty factor of 1.26--1.63 over the temperature range $T_9 = 0.2$--10. At higher temperatures, the propagated uncertainty in the ground-state contribution remains small, demonstrating that the present laboratory measurement provides the most robust experimental constraint to date on a substantial fraction of the stellar rate. A comparison with previously available experimental data shows overall consistency, while the recommended stellar rate is lower than the experimental constraint from Bhathi \textit{et al.} and higher than that reported by Kim \textit{et al.}. Moreover, the reaction rate derived here is lower than the REACLIB evaluation for $T_9 \leq 3$ and roughly consistent with the STARLIB rate.

Previous constraints on the $^{59}$Cu$(p,\gamma)^{60}$Zn reaction were reviewed. Despite remaining discrepancies in the $(p,\gamma)$ rate, the various evaluations are consistent to within a factor of 2 for temperatures $T_9 \gtrsim 0.3$. A direct comparison up to $T_9 \approx 3$ shows that the $(p,\alpha)$ rate remains consistently below the $(p,\gamma)$ rate across the relevant temperature range. This indicates that the NiCu cycle strength is small for temperatures relevant for XRBs.

While the present measurement, in combination with the earlier data of
Randhawa~\textit{et al.}~\cite{Jaspreet59Cu} and
Bhathi~\textit{et al.}~\cite{Bhathi25}, removes the $(p,\alpha)$ rate as
a significant source of uncertainty in the NiCu cycle branching over the
temperatures relevant to XRBs and the $\nu p$-process, the
$^{59}$Cu$(p,\gamma)^{60}$Zn rate now constitutes the dominant remaining
uncertainty.

\begin{acknowledgments}
This work is  based upon work supported by the U.S. Department of Energy, Office of Science, Office of Nuclear Physics, under Contract No. DE-AC02-06CH11357 and DE-AC05-00OR22725. It used resources of the Facility for Rare Isotope Beams (FRIB) Operations, which is a DOE Office of Science User Facility under Award Number DE-SC0023633. This work was also supported by the Institute for Basic Science (IBS) funded by the Ministry of Science and ICT, Korea (Grant No. IBS-R031-D1), and by NKFIH (K134197). This work was also performed under the auspices of the U.S. Department of Energy by Lawrence Livermore National Laboratory under contract DE-AC52-07NA27344.
\end{acknowledgments}

\bibliography{59Cu_pa}

\end{document}